\documentclass{JHEP3} 

\usepackage{epsfig}
\usepackage{amsmath, amssymb}
\usepackage{concmath, palatino}

\newcommand\fverb{\setbox\pippobox=\hbox\bgroup\verb}
\newcommand\fverbdo{\egroup\medskip\noindent%
			\fbox{\unhbox\pippobox}\ }
\newcommand\fverbit{\egroup\item[\fbox{\unhbox\pippobox}]}
\newbox\pippobox

\newcommand{\be}{\begin{equation}}
\newcommand{\ee}{\end{equation}}
\newcommand{\ba}{\begin{eqnarray}}
\newcommand{\ea}{\end{eqnarray}}

\newcommand{\pint}{\makebox[0pt][l]{\hspace{2.4pt}$-$}\int}

\newcommand{\la}{\longrightarrow}
\newcommand{\blt}{\ \ $\bullet$\ \ }

\newcommand{\ads}{AdS_5\times S^5}

\allowdisplaybreaks

\title{A numerical test of the Y-system in the small size limit of the $SU(2)\times SU(2)$ Principal Chiral Model}

\author{Matteo Beccaria and Guido Macorini\\
  Dipartimento di Fisica, Universita' del Salento, 
  Via Arnesano, 73100 Lecce\\
  INFN, Sezione di Lecce\\
  E-mail: \email{matteo.beccaria@le.infn.it}, \email{guido.macorini@le.infn.it} 
}

\abstract{
Recently,  Kazakov, Gromov and Vieira applied the discrete Hirota dynamics to study the finite size spectra
of integrable two dimensional quantum field theories. The method has been tested from large values of the 
size $L$ down to moderate values using the $SU(2)\times SU(2)$ principal chiral model as a theoretical laboratory. 
We continue the numerical analysis of the proposed non-linear integral equations showing that 
the deep ultraviolet region $L\to 0$ is numerically accessible. To this aim, we introduce a relaxed iterative  
algorithm for the numerical  computation 
of the low-lying part of the spectrum in the $U(1)$ sector. We discuss in details the systematic errors involved in the
computation. When a comparison is possible, full agreement is found with previous
TBA computations.}

\begin{document} 

\section{Introduction}
\label{sec:intro}

Finite size corrections to the complete spectrum of quantum field theories are an important issue with many 
applications. 
In particular, they are relevant to the aim of testing
Maldacena's AdS/CFT duality relating type IIB superstring propagating in $\ads$ and ${\cal N}=4$ SYM~\cite{Duality}.
In the 't Hooft planar limit, the gauge theory composite operators have anomalous dimensions related to the spectrum of 
a finite size integrable super spin chain~\cite{JPA}. The so-called {\em wrapping} corrections are crucial to check the 
correspondence and have been the subject of  major investigations in recent years~\cite{Wrapping}. 

\medskip
As shown by L\"uscher \cite{Luscher}, the leading corrections to the mass gap in rather general 
relativistic models
can be computed in terms of their infinite size scattering matrix. In integrable models, the $S$-matrix is 
one of the central objects and is known exactly. This suggests that the complete size corrections 
can be in reach.
Indeed, for continuum models admitting an integrable discretization, it is possible to determine the spectrum of all states
by solving the associated (nested) Bethe Ansatz equations. Their continuum limit leads to the  Destri-de Vega non-linear integral equations (DdV NLIE) \cite{ddv}. When an integrable discretization is not available, it is possible to 
determine the finite size
correction to the ground state energy by means of the thermodynamical 
Bethe Ansatz (TBA)~\cite{TBA,Hegedus:2004xd,Balog:2003yr}. The general idea is that,
in relativistic (Euclidean) 2d models, the free energy is related to the ground state energy in finite volume by a modular transformation exchanging spatial extension and inverse temperature.
Usually, the method leads to an infinite  system of 
coupled non linear integral equations which can be rewritten in the functional  Y-system 
form which is universal, {\em i.e.} symmetry based \cite{Ysystem}.

\medskip
Apparently, the very nature of the TBA construction forbids the computation of excited energies.
Nevertheless, various extensions have been proposed \cite{TBA}. Indeed, 
there are cases under full control, like for instance the Sine-Gordon 
model \cite{Balog:2003yr}, where the very same Y-system form of the TBA describes  
excited states. They are associated to solutions of the Y-system with 
different boundary conditions and analytical properties. 
Due to this important remark, the Y-system is the most promising object for the study of the full spectrum.

\medskip
In the AdS/CFT case, duality with string theory plays a crucial role by mapping the discrete spin chain describing
finite length operators to states of a continuum $\sigma$-model. In the end, it allows to apply the general idea of TBA 
and to obtain the remarkable formulations described in \cite{AdS-Y}~\footnote{Actually, the string is quantized in light-cone gauge which 
means that relativistic invariance is broken. This requires to compute the free energy of a mirror model which is 
not the same as the original theory.}. 
The resulting Y-system leads to  a very involved set of coupled non-linear integral equations where 
numerical methods are welcome, if not mandatory.

\medskip
Here, we begin a systematic investigation of such  numerical algorithms by going back to the initial analysis of Kazakov, Gromov
and Vieira who choose the Principal Chiral Model (PCM) as a theoretical laboratory.
In \cite{K1},  the authors (GKV) assume that the Y-system provides a complete description of the full spectrum
provided suitable solutions are considered. GKV describe a new systematic way to deal with the infinite 
component Y-system. 
They stress the important fact that the Y-system if closely related to the Hirota bilinear equation \cite{Ysystem}.
For finite rank symmetry group of the integrable model, the Hirota dynamics can be solved in terms of a finite number of 
functions of the spectral parameter.
Thus, the Y-system is successfully  reduced to a finite system of non-linear integral equations (NLIE) rather suitable for numerical methods.  

\medskip
In the GKV analysis, the proposed method is partially tested on the $SU(2)\times SU(2)$ Principal Chiral Model (PCM), equivalent to 
the $O(4)$ $\sigma$-model. Measuring all dimensional quantities in terms of the mass gap, the only parameter
which describes the spectrum is the continuous system size $L$. The low-lying part of the spectrum (including the 
ground state energy) is computed for $10^{-1}\le L\le 2$ by a numerical iterative algorithm. 
The general strategy starts from an approximate solution to the NLIE at large $L$. Then, the numerical solution is
found by iteration of the NLIE and the size $L$ is progressively reduced down to the desired value.
The ground state
and the first excited state energies can be compared with \cite{Hegedus:2004xd} where they are computed 
by means of 
coupled DdV-like equations, apparently unrelated to the GKV NLIE~\footnote{Although, {\em a priori}, 
it is not excluded that the two formulations can be related by some change of variable.}. The agreement is good and reasonably compatible with the expected behaviour in the deep ultraviolet region $L\ll 1$ which however remains  unexplored
at the smallest considered size, $L=0.1$.
In a subsequent paper, the analysis of \cite{K1} has been extended to the $SU(N)\times SU(N)$ PCM~\cite{K2}.
For $N=3$,  the spectrum of various excitations is extracted  as a function of the size $L$
down to $L=10^{-3}$. Unfortunately, for $N=4$ already $L\sim 1$ is not accessible due to instabilities in the 
iterative solution of the NLIE.

\medskip
In this paper, we reconsider the GKV integral equations for the $N=2$ model going down to very small 
size values providing a definitive test of their validity, at least in this region. 
We accomplish this task by a numerical analysis and present some refinements of the numerical strategy which,
in principle, 
could be useful for the study of $N>2$ models. In particular we adopt a straight discretization of the NLIE
which allows for a very clean control of the systematic errors. Also, we improve the iteration procedure by 
introducing various relaxation parameters turning out to be crucial in order  to reach $L=10^{-8}$.
These very small values of the size are even smaller than those achieved in previous studies of the deep UV region where
a comparison with asymptotically free field theory is possible \cite{Hegedus:2004xd,Balog:2003yr,Balog:2005yz}.
The outcome of our analysis is a very good agreement between the GKV equations and other known results. 
Besides, our numerical implementation is very simple and easily under control.

\medskip
The plan of the paper is the following. In Sec.~(\ref{sec:equations}), we briefly summarize the GKV approach and
its associated NLIEs. In Sec.~(\ref{sec:num-equations}), we describe our numerical algorithms for the ground state
and two excited states of interest. Finally, in  Sec.~(\ref{sec:results}), we discuss the numerical results
with particular attention to the role of systematic errors and  relaxation parameters. 

\section{Y-system equations}
\label{sec:equations}

In this section we briefly summarize the NLIE that we are going  to solve numerically. They are derived in full
details in \cite{K1}. Here we just sketch the main ideas and introduce the necessary quantities.

\medskip
The $SU(2)\times SU(2)$ Principal Chiral Model, equivalent to the $O(4)$ $\sigma$-model, is described by the action 
\be
S= -\frac{1}{2e_{0}^{2}}\,\int d^{2}x\,\mbox{Tr}\left(h^{-1}\partial_a h\right)^{2}, \qquad h\in SU(2).
\ee
This model is asymptotically free. The infinite size spectrum contains a single physical particle
with a dinamically generated mass $m = \Lambda_{UV}\,e^{\frac{-2\pi}{e_{0}^{2}}}$,  $\Lambda_{UV}$
being the ultraviolet cut-off. The associated state transforms in the bifundamental of   $SU(2)\times SU(2)$.
The elastic scattering matrix is known and can be written in terms of the scalar factor~\cite{Zamolodchikov:1978xm}
\be
S_{0}(\theta) = i\,\frac{\Gamma\left(\frac{1}{2}-\frac{i\theta}{2}\right)\,\Gamma\left(\frac{i\theta}{2}\right)}
{\Gamma\left(\frac{1}{2}+\frac{i\theta}{2}\right)\,\Gamma\left(-\frac{i\theta}{2}\right)}.
\ee
In general, one can consider states with $N$ particles and polarization defined by the number of left and right spins down~\footnote{
Here, left and right refer to the two $SU(2)$ factors, or wings.} .
In this paper we shall consider the so-called $U(1)$ sector
where all spins are up (ferromagnetic polarization). The infinite size energy of such states can be computed 
from the formula
\be
E = m\,\sum_{i=1}^{N}\cosh(\pi\,\theta_{i}),
\ee
where the physical rapidities $\{\theta_{i}\}_{i=1,\dots, N}$ solve 
the asymptotic Bethe Ansatz~\cite{Gromov:2006dh}
\be
e^{-i\,L\,\sinh(\pi\theta_{i})} = -\prod_{j=1}^{N} S_{0}^{2}(\theta_{i}-\theta_{j}) .
\ee
In the TBA approach, the finite size ground state energy is computed starting from the infinite size
thermodynamics associated to the  above equations. If we denote by $L$ the size in units of the mass gap, then the result is the following simple formula for the energy
\be
\label{eq:gsenergy}
E_{0}(L) = -\frac{1}{2}\int_{\mathbb R} d\theta \cosh(\pi\,\theta)\,\log(1+Y_{0}),
\ee
where $Y_{0}$ is obtained from the solution of the TBA equations. These are integral equations which, upon reasonable physical assumptions, can be written in the celebrated $Y$-system form~\footnote{
The notation is $f^{{\pm\dots\pm}}(x) = f(x\pm \frac{i\,p}{2})$, when we have $p$ plus or minus signs.}
\be
\label{eq:Ysystem}
Y^{+}_{n}Y^{-}_{n} = (1+Y_{n-1})(1+Y_{n-1}), \qquad n\in\mathbb{Z}.
\ee
The physical boundary conditions predict the large $\theta$ behaviour of $Y_{n}$. All $Y_{n}$ with $n\neq 0$
tend to a constant,  while $Y_{0}$ is exponentially suppressed according to 
\be
Y_{0}(\theta)\sim e^{-L\,\cosh(\pi\theta)}.
\ee
As we explained in the introduction, the central problem is that of solving the $Y$-system since the main assumption
is that it describes all states and not only the ground state. 

\medskip
The GKV method to study Eq.~(\ref{eq:Ysystem}) is based on the relation to the Hirota equation that
 can be written as ($\overline\Phi$ is the conjugate of $\Phi$)
\be
\label{eq:Hirota}
T_{n}\left(\theta+\frac{i}{2}\right)\,T_{n}\left(\theta-\frac{i}{2}\right)
-T_{n-1}(\theta)\,T_{n+1}(\theta) = \Phi\left(\theta+\frac{in}{2}\right)\, \overline\Phi\left(\theta-\frac{in}{2}\right),
\ee
where the functions $T_{n}$ are related to $Y_{n}$ by the relation
\be
\label{eq:YT}
Y_{n}(\theta) = \frac{T_{n+1}(\theta)T_{n-1}(\theta)}{\Phi\left(\theta+\frac{in}{2}\right)\overline
\Phi\left(\theta-\frac{in}{2}\right)} = \frac{T_{n}\left(\theta+\frac{i}{2}\right)T_{n}\left(\theta-\frac{i}{2}\right)}
{\Phi\left(\theta+\frac{in}{2}\right)\overline
\Phi\left(\theta-\frac{in}{2}\right)}-1.
\ee
Eqs.~(\ref{eq:Hirota}, \ref{eq:YT}) are gauge invariant according to 
\ba
Y_{n}(\theta) &\la& Y_{n}(\theta), \nonumber \\
T_{n}(\theta) &\la& g\left(\theta+\frac{in}{2}\right)\,\overline g\left(\theta-\frac{in}{2}\right)\,T_{n}(\theta),
\\
\Phi(\theta) &\la& g\left(\theta-\frac{i}{2}\right)\, g\left(\theta+\frac{i}{2}\right)\,\Phi(\theta).
\nonumber 
\ea 
The Hirota equation is integrable (it admits a discrete Lax pair) and this allows for a complete analysis of its 
solutions \cite{K1}.  

\medskip
The main trick of GKV is a clever continuation from the case of large $L$ where a fundamental important simplification occurs. Indeed,  the $L\to \infty$ suppression of $Y_{0}$
leads to a decoupling of the $Y$-system in the two sets $\{Y_{n}\}$ with $n>0$ or $n<0$. Both can be written, 
according to (\ref{eq:YT}), in terms of a suitable solution of the Hirota equation and thus we identify two independent solutions of it. This picture can be continued to finite $L$ as carefully shown in \cite{K1}. The resulting 
two independent solutions are no more decoupled and give nontrivial $T_{n}$ for all $n$ on both wings of the 
$SU(2)\times SU(2)$ Dynkin diagram. These two solutions describe the same state and must be related by a gauge 
tranformation $g(\theta)$. This consistency condition gives a single NLIE for the function $g$ which is precisely the equation we are going to study. Once this equation is solved, the function $Y_{0}$
associated to the desired excited states is available, and the formula for the energy simply becomes
\be
\label{eq:excited}
E(L) = m\,\sum_{i=1}^{N}\cosh(\pi\,\theta_{i})-\frac{1}{2}\int_{\mathbb R} d\theta \cosh(\pi\,\theta)\,\log(1+Y_{0}).
\ee

\medskip
Let us introduce the definitions
\be
A(\theta) = (g^{+}(\theta))^{2}, \qquad K_{0}(\theta) = \frac{1}{2\,\pi\,i}\frac{d}{d\theta}\,\log S_{0}^{2}(\theta).
\ee
The function $A(\theta)$ is determined by the NLIE \cite{K1} (a star denotes as usual convolution with respect to the
rapidity argument)
\be
\label{eq:GKV1}
A = -e^{-L\,\cosh(\pi\,\theta)}\prod_{i=1}^{N}S_{0}^{2}\left(\theta-\theta_{i}+\frac{i}{2}\right)\,
\exp\left(
K_{0}*\log\frac{A-1}{|A|-1}-K_{0}^{++}*\log\frac{\overline A-1}{|A|-1}
-\log\frac{\overline A-1}{|A|-1}
\right),
\ee
where the physical rapidities are self consistently determined by 
\be
\label{eq:GKV2}
-e^{i\,L\,\sinh(\pi\,\theta_{i})}\prod_{j=1}^{N}S_{0}^{2}(\theta_{i}-\theta_{j})
\,\exp\left[
2i\,\mbox{Im}\left(
K_{0}^{-}*\log\frac{A-1}{|A|-1}
\right)
\right]_{\theta=\theta_{i}} = 1.
\ee
According to Eq.~(\ref{eq:excited}), the energy is given by 
\be
E(L) = \sum_{i=1}^{N}\cosh(\pi\theta_{i})-\int_{\mathbb{R}} d\theta \,\cosh(\pi\theta)\,\log\frac{A-1}{|A|-1}.
\ee
This formula can also be used for the ground state which is associated with $N=0$. In this case, all factors 
$S_{0}$ must be replaced by 1, leaving of course the kernel $K_{0}$ untouched. 

\subsection{A second formula for the ground state energy}

Remarkably, an alternative NLIE for the ground state energy is also presented in \cite{K1}. It is a self-consistent  equation
for the function $f(\theta)$ which  reads~\footnote{
Note that the singular convolution is more explicitly
\be
\frac{1}{\pi}\int_{\mathbb{R}} d\xi \frac{1}{4\,(\theta-\xi+\frac{i}{2}+i\,0)^{2}+1}\,f(\xi) = 
\frac{1}{\pi}\pint_{\mathbb{R}} d\xi \frac{1}{4\,(\theta-\xi+\frac{i}{2})^{2}+1}\,f(\xi) -\frac{1}{4}\,f(\theta).
\ee
}
\be
\label{eq:secondmethod}
f(\theta) = 2\,T_{1}(\theta)\,\left|T_{0}\left(\theta+\frac{i}{2}+i\,0\right)\right|^{2}\,
\exp\left[
-L\cosh(\pi\theta)-2s*\log|T_{0}^{++}|^{2}
\right],
\ee
where $T_{n}$ are obtained from $f$ according to
\be
T_{n-1} = n+\frac{n}{\pi\,(4\theta^{2}+n^{2})} * f,
\ee
and where the function $s(\theta)$ is 
\be
s(\theta) = \frac{1}{2\,\cosh(\pi\theta)}.
\ee
Once Eq.~(\ref{eq:secondmethod}) is solved, the energy is given by (\ref{eq:gsenergy}) with the following expressions for $Y_{0}$
\be
Y_{0} = \frac{T_{1}(\theta)\,f(\theta)}{2\,\left|T_{0}\left(\theta+\frac{i}{2}+i\,0\right)\right|^{2}}
\ee
We shall discuss the numerical accuracy and efficiency of this second approach comparing it with the 
general formula for $N$ particle states specialized at $N=0$.

\section{Numerical implementation}
\label{sec:num-equations}

\subsection{Discretization}

A numerical implementation of Eqs.~(\ref{eq:GKV1}, \ref{eq:GKV2}) has been presented in \cite{K1} in terms 
of a short Mathematica code exploiting various high level functions for numerical integration and interpolation. 
Instead, 
our approach will be that of presenting a simple algorithm which can kept easily under control in order to check the 
continuum limit and the bias due to systematic errors. To this aim we cut-off the rapidity $\theta$ in the interval 
$|\theta|\le \Lambda$ and split it introducing the discrete points
\be
\theta_{n} = \Lambda\,\left(-1+2\,\frac{n-1}{M-1}\right) \equiv \Lambda\,\xi_{n}, \qquad n=1, \dots, M.
\ee
For numerical efficiency, we define the fixed vectors and matrices
\ba
Z^{(1)}_{n} &=& -e^{-L\cosh(\pi\theta_{n})}\,S_{0}\left(\theta_{n}+\frac{i}{2}\right)^{2}, \qquad n=1, \dots, M, \nonumber \\
\widetilde Z^{(2)}_{n} &=& \left\{
\begin{array}{ll}
\displaystyle \frac{2\log 2}{\pi}, & n=0, \\
\displaystyle K_{0}\left(\frac{2\,n\,\Lambda}{M-2}\right) & n\neq 0, 
\end{array}
\right. \qquad n=-M, \dots, M, \\
\widetilde Z^{(3)}_{n} &=& \left\{
\begin{array}{ll}
\displaystyle 0, & n=0, \\
\displaystyle K_{0}\left(\frac{2\,n\,\Lambda}{M-2}+i\right) & n\neq 0, 
\end{array}
\right. \qquad n=-M, \dots, M, \nonumber \\
Z^{(2)}_{nm} &=& \widetilde Z^{(2)}_{n-m+M+1}, \qquad \qquad n, m =1, \dots, M, \nonumber \\
Z^{(3)}_{nm} &=& \widetilde Z^{(3)}_{n-m+M+1},  \qquad \qquad n, m =1, \dots, M, \nonumber 
\ea
They are evaluated with a (large) fixed number of digits of order 200. The issue of numerical precision will
be discussed later. We now present the numerical algorithms for the ground state, the first excited state which 
has a zero Bethe root, and
another state which is non trivial from the Bethe Ansatz point of view having two opposite Bethe roots.

\subsection{Fixed point search}
\label{sec:fp}

The NLIE that we want to solve takes the form of a fixed point equation $f(x)=x$.
It can be solved by iteration when the map $x\mapsto f(x)$ is a contraction. In other words,
one simply sets $x_{n+1}=f(x_{n})$ and the sequence $\{x_{n}\}$ converges to the solution $x^{*}$ as $n\to \infty$,
independent on the starting point. In actual cases, one needs to start in a suitable small neighborhood of $x^{*}$. Then, 
the map can be linearized in terms of the small differences $\delta_{n} = x_{n}-x^{*}$ and one has to iterate
\be
\delta_{n+1} = a\,\delta_{n},\qquad a \equiv f'(x^{*}).
\ee
Convergence is exponentially good for $|a|<1$, otherwise the sequence diverges. Following Jacobi and Seidel, 
a simple general trick to deal with the 
case $|a|>1$ is to introduce a relaxation parameter 
and consider some average between the old and new values
of the $\delta$ variables. In other words, one considers the following modified equation 
\be
x_{n+1} = (1-\lambda)\,x_{n}+\lambda\,f(x_{n}),
\ee
where $\lambda$ is a real number. The advantage is clear. Around the fixed point one has
\be
\label{eq:linearized}
\delta_{n+1} = (1-\lambda+\lambda\,a)\,\delta_{n}.
\ee
This means that in the unstable case, $|a| >1$, we can enforce stability by simply choosing
\be
\label{eq:relax}
\left\{
\begin{array}{lcl}
a>1 &:&  \displaystyle -\frac{2}{a-1} < \lambda < 0, \\
a<-1 &:& \displaystyle \ \  \ \quad \quad 0<\lambda<\frac{2}{1-a}, \\
\end{array}
\right.
\ee
In the $p$-component case, where $x_{n}\in \mathbb{C}^{p}$, the same argument based on the linearization around
the fixed point, suggest to replace $\lambda$ by a $p\times p$ matrix and to apply (\ref{eq:relax}) to each eigenvalue
of the matrix $\nabla f$. More involved schemes can be devised replacing  (\ref{eq:linearized}) by higher order 
relaxation processes as in the inertial relaxation discussed in \cite{FP1} or also finding adaptive algorithms for the step by step choice of $\lambda$ \cite{FP2}. In this paper, we shall discuss the simplest implementation of the relaxation idea
as illustrated above.

\subsection{Algorithm for the excited state $\Theta_{0}$}
\label{sec:theta0}

Let us begin our discussion with the 1-particle state $\Theta_{0}$. It is
discussed in \cite{K1} and is the lightest excited state. So, it enters the mass gap computation.
By symmetry, there is only one
Bethe roots which is fixed at zero. This simplifies a lot the computation.

The function $A(\theta)$ becomes a sequence of vectors $\{A^{(k)}_{n}\}_{k\ge 0, 1\le n\le M}$ which takes the initial value
\be
A^{(0)}_{n} = Z_{n}^{(1)} = -e^{-L\cosh(\pi\theta_{n})}\,S_{0}\left(\theta_{n}+\frac{i}{2}\right)^{2}.
\ee
Then, the main loop starts.  Defining 
\be
r^{(k)}_{n} = \log\frac{A^{(k)}_{n}-1}{|A^{(k)}_{n}|-1},
\ee
we have
\ba
A_{n}^{(k)} &=& (1-\lambda)\,A_{n}^{(k-1)}+ \\
&& +\lambda\,Z^{(1)}_{n}\,
\exp\left[
\frac{2\Lambda}{M-2}\sum_{m=1}^{N}\,w_{m}\,\left(
Z^{(2)}_{nm}\,r^{(k-1)}_{m}-Z^{(3)}_{nm} \overline r^{(k-1)}_{m}\right)
-\overline r^{(k-1)}_{n}\right].\nonumber 
\ea
The parameter $\lambda$ is a relaxation parameter introduced according to the previous discussion.
It is  crucial at small $L$ as we shall discuss. The algorithm in \cite{K1}
is obtained for $\lambda=1$ and works for $L$ of order $10^{-1}$. We shall push the computation several 
orders of magnitude beyond this value. In this deep UV region, the above map is no more a contraction and 
convergence is lost. Nevertheless, a moderate value of $\lambda\in (0,1)$~\footnote{We never observed the necessity of over-relaxation.} will turn out to be enough to 
restore convergence. The weights $w_{m}$ are used to approximate integrations by summations. They can be 
taken at the rough value $w_{m}\equiv 1$ or improved as in Simpson or higher order integrations. For our
purposes this is not a crucial issue as we shall explain later.

The very same algorithm can be applied to the ground state. The only necessary change is the replacement of 
all $S_{0}$ factors by unity. In conclusion the parameters entering the numerical algorithm are
\be
\Lambda, \qquad M, \qquad \lambda.
\ee

\subsection{Algorithm for the excited state $\Theta_{00}$}

This is a 2-particle state. By symmetry, it is associated with a pair of opposite rapidities $\pm\theta^{*}$
solving the Bethe Ansatz equations. Again, we start from the method presented in \cite{K1}, but add the important
relaxation parameters. Indeed, it is convenient to choose two independent relaxation times for the 
function $A$ and for the Bethe root(s). This is a fundamental point otherwise convergence is soon lost.

Let us denote by $\{\widetilde\theta_{a}^{(k)}\}_{a=1,2}$, with $\widetilde\theta_{2}^{(k)}=
-\widetilde\theta_{1}^{(k)}$, the two Bethe rapidities. 
They satisfy
the BA equations (actually it is one equation due to the 
parity symmetry !) \cite{K1},
\be
L\sinh(\pi\widetilde\theta^{(k)}_{a})-i\mathop{\sum_{a\neq b}^{2}}_{b=1}
\log S_{0}(\widetilde\theta^{(k)}_{a}-\widetilde\theta^{(k)}_{b})^{2}+\frac{4\Lambda}{M-2}
\sum_{\ell=1}^{M} \mbox{Im}\left(
K_{0}\left(
\widetilde\theta^{(k-1)}_{a}-\theta_{\ell}-\frac{i}{2}
\right)\,r^{(k-1)}_{\ell}
\right) = 0.
\ee
The initial value of the Bethe roots is not critical and a possible choice is just to solve the above equation
suppressing the last term in the left hand side. The modified main loop is then 
\ba
A_{n}^{(k)} &=& (1-\lambda_{1})\,A_{n}^{(k-1)}+ \\
&& +\lambda_{1}\,Z^{(1)}_{n}\,\prod_{a=1}^{2}S_{0}(\theta_{n}-\widetilde\theta_a^{(k-1)}+\frac{i}{2})^{2}
\times \nonumber \\
&&\times \exp\left[
\frac{2\Lambda}{M-2}\sum_{m=1}^{N}\,w_{m}\,\left(
Z^{(2)}_{nm}\,r^{(k-1)}_{m}-Z^{(3)}_{nm} \overline r^{(k-1)}_{m}\right)
-\overline r^{(k-1)}_{n}\right].\nonumber 
\ea
In addition, we must provide an update rule for the Bethe roots. We propose 
\be
\widetilde\theta^{(k)}_{a} = (1-\lambda_{2})\,\widetilde\theta^{(k-1)}_{a} +\lambda_{2}\, \widetilde\theta^{(k-1)}_{a}.
\ee
In summary, the parameters entering the numerical algorithm are
\be
\Lambda, \qquad M, \qquad \lambda_{1}, \qquad \lambda_{2}.
\ee
As a remark, we note that for more complicated states with more Bethe roots, we still have two relaxation parameters,
one for the function $A$ and one for the Bethe roots.

\subsection{Alternative algorithm for the ground state}

The numerical implementation of the solution of  (\ref{eq:secondmethod}) is completely analogous
to that described in Sec.~(\ref{sec:theta0}). The initial value of the discretized $f$ function is 
\be
f^{(0)}_{n} =4\,e^{-L\cosh(\pi\theta_{n})}.
\ee

\section{Numerical Results}
\label{sec:results}

We present here our numerical results aimed at understanding (a) the systematic errors associated with finite values of 
$\Lambda$, $M$, (b)  the role of the relaxation parameters $\lambda_{i}$.
Let us begin with the cut-off $\Lambda$. In principle, one should consider a fixed $\Lambda$ and send $M\to \infty$
in order to reach the continuum limit.
The result is a function of $\Lambda$ which has to be extrapolated at $\Lambda\to \infty$. In practice, as soon as 
$\Lambda$ is large enough to cover the support of the function $A(\theta)$, one observes independence on $\Lambda$
with exponentially increasing accuracy. In the following we shall set 
\be
\Lambda = \frac{n_{\Lambda}}{\pi}\,\mbox{arccosh}\left(\frac{8\log 10}{L}\right).
\ee
When $n_{\Lambda}=1$, we have the exponential factor $e^{-L\,\cosh(\pi\,\Lambda)} = 10^{-8}$ which is a reasonable 
definition of a {\em small} quantity in this problem. We shall work with several hundreds of digits which will be 
enough in all cases where convergence is achieved.

\subsection{Ground State}

We compared the computation of the ground state energy based on (\ref{eq:GKV1}) or (\ref{eq:secondmethod}).
The first equation provides much more accurate results and shall be discussed in this section. In the final summary table we shall present results for the second method too.
As a preliminary step, we start with $L=10^{-1}$ which is the smallest value considered
in \cite{K1}. We present the time history of the energy iterates at relaxation $\lambda = \frac{1}{2}$ in 
Fig.~(\ref{fig:GS-1-relaxation}). The discretization is $M=350$. One sees that there is easy convergence 
to a value which has a mild residual dependence on $\Lambda$. This dependence can be observed together with the 
dependence on $M$ in Fig.~(\ref{fig:GS-1-lines}). A natural simple guess is to assume that for large enough $\Lambda$
(all considered cases should be all right from this point of view), the relevant parameter is the density of points $\Lambda/M$. This hypothesis is tested in Fig.~(\ref{fig:GS-1-density}). One sees that indeed the dependence on $\Lambda$ and $M$ is under control. The considered values of $\Lambda$ are already asymptotic and the dependence
on $M$ is quite accurately linear~\footnote{
In principle one can improve the convolutions by improved numerical integrations. Actually, this would be an 
incomplete improvement and we have checked that residual linear corrections $1/M$ do remain. 
}. As an interesting additional information,
we provide in Fig.~(\ref{fig:GS-1-profile}) the real and imaginary parts of the function $A(\theta)$. One sees the initial 
profile before the iteration loop as well as the final equilibrium value.

The same analysis can be repeated at the very smaller value 
$L=10^{-6}$. Again, we present the time history of the energy iterates at relaxation $\lambda = \frac{1}{2}$ in 
Fig.~(\ref{fig:GS-2-relaxation}). The discretization is $M=500$. The dependence on $\Lambda$ and $M$
is shown in Fig.~(\ref{fig:GS-2-lines}). The scaling plot showing dependence on $\Lambda/M$
is Fig.~(\ref{fig:GS-2-density}). Finally, we provide in Fig.~(\ref{fig:GS-2-profile}) the real and imaginary parts of the function $A(\theta)$. 

This small value of $L$ permits to emphasize the role of relaxation. This is shown in Fig.~(\ref{fig:GS-2-lambda})
where we show the time history of the energy as the relaxation parameter is reduced from 1, where instability is 
observed,  to $1/3$ where convergence is all right.

\subsection{First excited state $\Theta_{0}$}

We present our numerical data for the excited state $\Theta_{0}$ following the same scheme
as for the ground state.
In particular, we show for $L=10^{-1}$ the time history of the energy iterates at relaxation $\lambda = \frac{1}{2}$ in 
Fig.~(\ref{fig:Theta0-1-relaxation}). The discretization is $M=350$. The dependence on $\Lambda$ and $M$
is shown in Fig.~(\ref{fig:Theta0-1-lines}). The scaling plot showing dependence on $\Lambda/M$
is Fig.~(\ref{fig:Theta0-1-density}). Similar plots for $L=10^{-6}$ can be found in 
Figs.~(\ref{fig:Theta0-2-relaxation}, \ref{fig:Theta0-2-lines}, \ref{fig:Theta0-2-density})).
Finally, we provide in Fig.~(\ref{fig:Theta0-2-profile}) the real and imaginary parts of the function $A(\theta)$. 

Again, at $L=10^{-6}$, we can show the role of relaxation. This is shown in Fig.~(\ref{fig:Theta0-2-lambda})
where we show the time history of the energy as the relaxation parameter is reduced from 1, where instability is 
observed,  to $1/2$ where convergence is all right.

\subsection{Excited state $\Theta_{00}$}

This is the next-to-lowest excited state, at least for small enough size $L$. The relaxation algorithm depends now
on two independent parameters: $\lambda_{1}$ for the update of the function $A$ and $\lambda_{2}$
for the update of the Bethe root(s). We present in Figs.~(\ref{fig:Theta00-2-energy-lambda},
\ref{fig:Theta00-2-theta-lambda}) the time history of the energy of the state $\Theta_{00}$
as well as that of $\widetilde\theta_{1}$ for $M=50$, $n_{\Lambda}=2$, and fixed $\lambda_{1}=1/2$.
As the second parameter $\lambda_{2}$ is reduced, we move from an oscillating regime to a convergent one.
The onset of oscillations instead of an exponential instability is compatible with the fact that we are dealing with a 
two component system.

Next, using the parameters shown in Tab.~(\ref{tab:theta00}), we present at $L=10^{-6}$ similar plots to those that we have discussed for the other states. In particular, 
 with discretization $M=600$, we show the dependence on $\Lambda$ and $M$
 in Fig.~(\ref{fig:Theta00-2-lines}), the scaling plot in Fig.~(\ref{fig:Theta00-2-density}), and 
 in Fig.~(\ref{fig:Theta00-2-profile}) the real and imaginary parts of the function $A(\theta)$. 

Finally, in Fig.~(\ref{fig:Theta00-all-profile}) we show the profile of $A(\theta)$ 
which is obtained after convergence at various sizes $L$. These could help, at least in principle, in the formulation
of suitable proposals for the analytic size dependence of this important function.

\subsection{Summary tables and $L\to 0$ limit}

Our final results, obtained after extrapolating to $M\to \infty$,  are summarized in three tables. For the ground state, they are shown in Tab.~(\ref{tab:gs}).
The first column reports the results obtained with the TBA NLIE of \cite{Hegedus:2004xd} down to $L=10^{-6}$.
The next column shows the results obtained using the equation (\ref{eq:secondmethod}). Finally, the third column
shows the results which we obtain using Eq.~(\ref{eq:GKV1}) which turns out to be much more 
efficient and accurate. The Y-system results have been obtained reducing the size by two orders of magnitude compared 
with \cite{Hegedus:2004xd}. The similar Tab.~(\ref{tab:theta0}) presents our results for the energy of the 
state $\Theta_{0}$ while we present in Tab.~(\ref{tab:theta00})  the energy of $\Theta_{00}$ for which 
there are no available results obtained with other methods.

One can check that there is a very good agreement showing that the GKV equations are working perfectly in the 
very small size limit. A better precision could be achieved by simply increasing $M$ in order to reduce the effect of the 
subleading correction. To the aim of testing the GKV equations, we honestly believe that the quality of our result is 
convincing.

In Fig.~(\ref{fig:final-plot}), we show the plot of the three energies as functions of $L$.
The mass gap is clearly reproduced given the agreement between Tab.~(\ref{tab:gs}) and the analysis
of \cite{Hegedus:2004xd}. There is an additional check that we can perform on our data, 
and that follows the analysis of \cite{K1}. Indeed, all the three considered energies should have the limit
$\frac{L}{2\,\pi}\,E\to -1/4$ as $L\to 0$. This limit is definitely out of reach for the numerics
of \cite{K1}. Fitting our numbers 
in the range  $10^{-8} \le L \le 10^{-5}$ by means of a (naive)
quadratic polynomial in $1/\log L$ we obtain the three estimates
\be
\mbox{ground\ state}\ :\ -0.249(3), \qquad
\Theta_{0}\ :\ -0.2477(1), \qquad
\Theta_{00}\ :\ -0.2439(3). 
\ee
These extrapolations are rather close to the predicted limit $-0.25$. 
The quoted errors are simply those
inherited from the data. There is no attempt to estimate the systematic error due to subleading corrections
in $L\to 0$ which are apparently larger for the excited states.

\begin{table}
\begin{center}
\begin{tabular}[t]{c||c|c|c|c|}
$L$ & $E_{0}$ (NLIE) \cite{Hegedus:2004xd}  & $E_{0}$ (Y-System)  Eq.~(\ref{eq:secondmethod}) & $E_{0}$ (Y-System) & Parameters \\
\hline
$10^{-1}$ & $ -11.273364587(1) $ &  $-11.273(1)$  & $ -11.2734(2) $ & $M=200-350$, $\lambda=1/2$ \\
\hline
$10^{-2}$ & $ -127.22634373(1) $ &   $-127.22(8) $	& $ -127.226(2)$ &  $M=250-400$, $\lambda=1/2$\\
\hline
$10^{-3}$ & $ -1343.4090793(1) $ &   $-1343(3)$	& $ -1343.46(2)$  &  $M=300-450 $, $\lambda=1/2$\\
\hline
$10^{-4}$ & $ -13865.238816(1) $ &   $-1.386(4) \times 10^{4}$	& $  -13865.8(2)$  &  $M=350-500 $, $\lambda=1/2$\\
\hline
$10^{-5}$ & $ -141563.8217(1) $  &    $-1.415(5) \times 10^{5}$	& $  -141564(5)$    &  $M=350-500 $, $\lambda=1/10$\\
\hline
$10^{-6}$ & $ -1436683.423(1) $  &    $-1.434(7) \times  10^6$  & $-1.4367(1)\times 10^6$ & $M=350-500 $, $\lambda=1/10$\\ 
\hline
$10^{-7}$ &                      &	                                   & $-1.4526(2) \times10^7 $ & $M=350-500 $, $\lambda=1/15$ \\
\hline
$10^{-8}$ &                      &	                                   &  $-1.4651(4) \times 10^8 $ & $M=350-500 $, $\lambda=1/15$ \\
\end{tabular}
\end{center}
\caption{Ground states energies. The first column is taken from \cite{Hegedus:2004xd}. The second column
reports our best results. The last column shows the range of discretization $M$ and the relaxation parameter.}
\label{tab:gs}
\end{table}

\begin{table}
\begin{center}
\begin{tabular}[t]{c||c|c|c|}
$L$ & $E(\Theta_0)$ (NLIE) \cite{Hegedus:2004xd}  &  $E(\Theta_0)$ (Y-System) &
Parameters \\
\hline
$10^{-1}$ & $-3.004108884(1)$ & $-3.004109(8)$ & $M=200-350$, $\lambda=1/2$\\
\hline
$10^{-2}$ & $-69.83802786(1)$ & $-69.8380(3)$ & $M=350-400$, $\lambda=1/2$\\
\hline
$10^{-3}$ & $-901.2815867(1)$ & $-901.282(9)$ & $M=400-450$, $\lambda=1/2$\\
\hline
$10^{-4}$ & $-10260.214298(1)$ & $-10260.2(1) $ & $M=450-500$, $\lambda=1/2$\\
\hline
$10^{-5}$ & $-111091.0324(1)$ & $-111091(2)$ & $M=500-550$, $\lambda=1/2$\\
\hline
$10^{-6}$ & $-1172575.496(1)$ & $-1.17258(2) \times 10^6$ & $M=550-600$, $\lambda=1/2$\\
\hline
$10^{-7}$ & & $-1.21947(1)  \times 10^7$ & $M=550-600$, $\lambda=1/2$\\
\hline
$10^{-8}$ &  & $-1.25637(1)  \times  10^8$ & $M=550-600$, $\lambda=1/2$\\
\end{tabular}
\end{center}
\caption{$\Theta_0$ energies. The first column is taken from \cite{Hegedus:2004xd}. The second column
reports our best results. The last column shows the range of discretization $M$ and the relaxation parameter.}
\label{tab:theta0}
\end{table}

\begin{table}
\begin{center}
\begin{tabular}[t]{c||c|c|}
$L$ &  $E(\Theta_{00})$ (Y-System) &
Parameters \\
\hline
$10^{-1}$ & $12.423491(8)$ & $M=200-350$,
$\lambda_1=1/2$, $\lambda_2=\lambda_1/5$  \\
\hline
$10^{-2}$ & $52.4704(3)$   & $M=200-350$, $\lambda_1=1/2$, $\lambda_2=\lambda_1/5 $\\
\hline
$10^{-3}$ & $123.324(3)$    & $M=250-400$, $\lambda_1=1/2$, $\lambda_2=\lambda_1/5 $\\
\hline
$10^{-4}$ & $-1515.85(2)$    & $M=250-400$, $\lambda_1=1/2$, $\lambda_2=\lambda_1/5 $\\
\hline
$10^{-5}$ & $-35284(1)$      & $M=300-450$, $\lambda_1=1/2$, $\lambda_2=\lambda_1/5 $\\
\hline
$10^{-6}$ & $-5.0557(1)\times 10^{5}$   & $M=300-450$, $\lambda_1=1/5$, $\lambda_2=\lambda_1/10 $\\
\hline
$10^{-7}$ & $  -6.2486(2) \times 10^6 $ & $M=200-450$,  $\lambda_1=1/5$, $\lambda_2=\lambda_1/20 $\\
\hline
$10^{-8}$ & $-7.2033(6) \times 10^7 $ & $M=200-350$,  $\lambda_1=1/5$, $\lambda_2=\lambda_1/20 $\\
\end{tabular}
\end{center}
\caption{$\Theta_{00}$ energies. The first  column
reports our best results. The second column shows the range of discretization $M$ and the relaxation parameters.}
\label{tab:theta00}
\end{table}

\section{Conclusions}

The recent proposal of GKV \cite{K1} provides a quite general method to compute finite size correction to 
the full spectrum of two dimensional integrable models. It involves non linear integral equations that
can be treated in the full space of physical parameters  only by means of numerical methods.
Whenever integrable discretizations are not available, the calculation of excited levels is based on 
certain assumptions. For this reason, it is important to compare different methods as well as achieve
accurate numerical predictions. In this paper we have worked out the small size limit of the $SU(2)\times SU(2)$
Principal Chiral
Model. To this aim, we have tested a numerical implementation of the NLIE of \cite{K1}. We have explored the possibility of solving them by iteration in the case of the ground state and 
of two additional excited states. We found that small $L$ values require 
to introduce relaxation constants in order to achieve convergence. This has to be done independently for the 
NLIE and Bethe root evolution.
We hope that this investigation will be useful in the analysis of the $SU(N)\times SU(N)$
Principal Chiral
Model for general $N$ along the lines of \cite{K2}. We believe that this first steps are necessary in order to attack
the full Y-system equations for the AdS/CFT problem \cite{AdS-Y} having all the systematic errors of the numerics
under control.




\newpage
\FIGURE{
\label{fig:GS-1-relaxation}
\epsfig{width=12cm,file=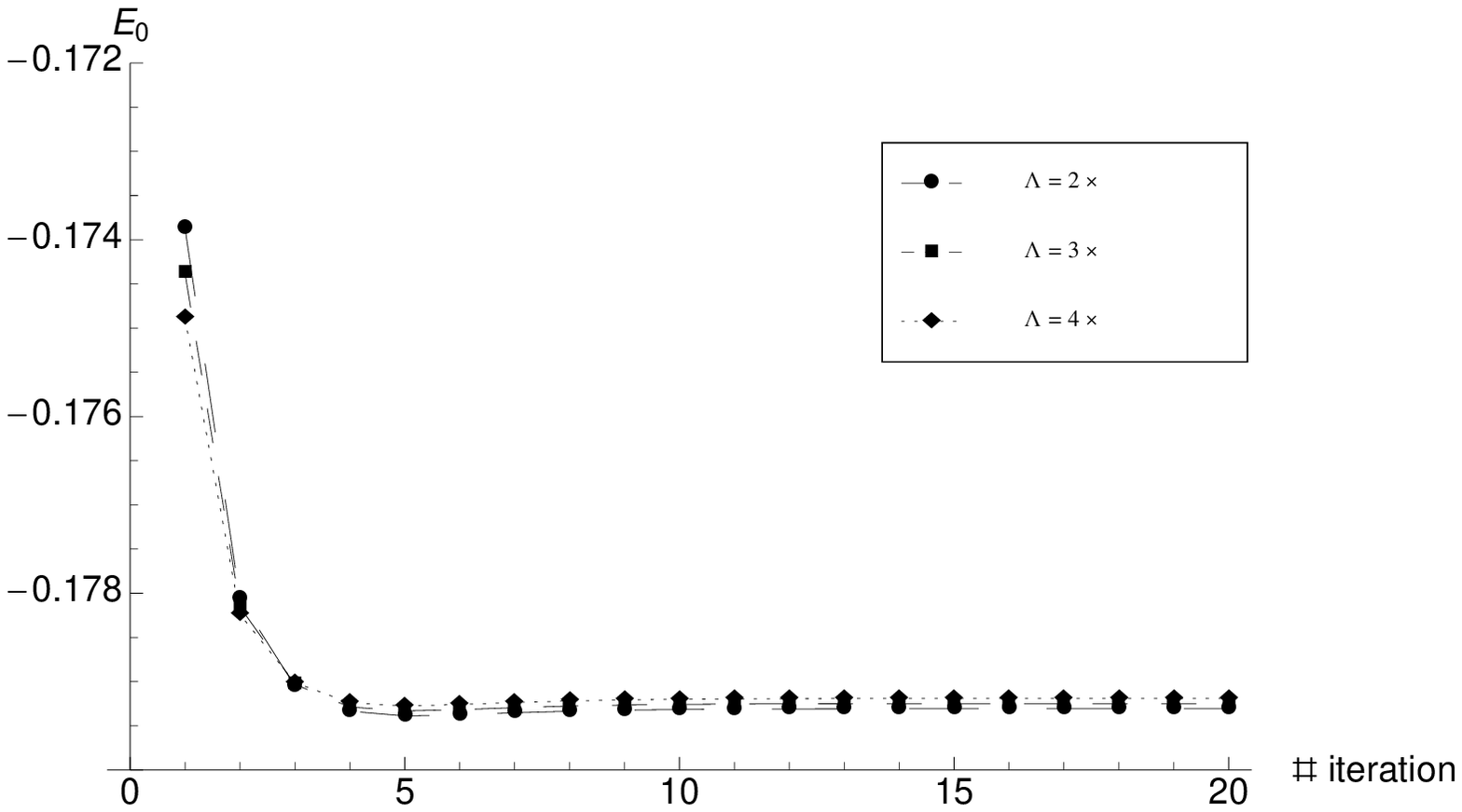}
\caption{
$L=10^{-1}$, $M=350$. Time history of the ground state energy iterates with relaxation $\lambda = 1/2$. The notation 
of the legend is $\Lambda = n_{\Lambda}\times \cdots$. The three values of $n_{\Lambda}$ are used to check
 cutoff independence.
}
}

\FIGURE{
\label{fig:GS-1-lines}
\epsfig{width=12cm,file=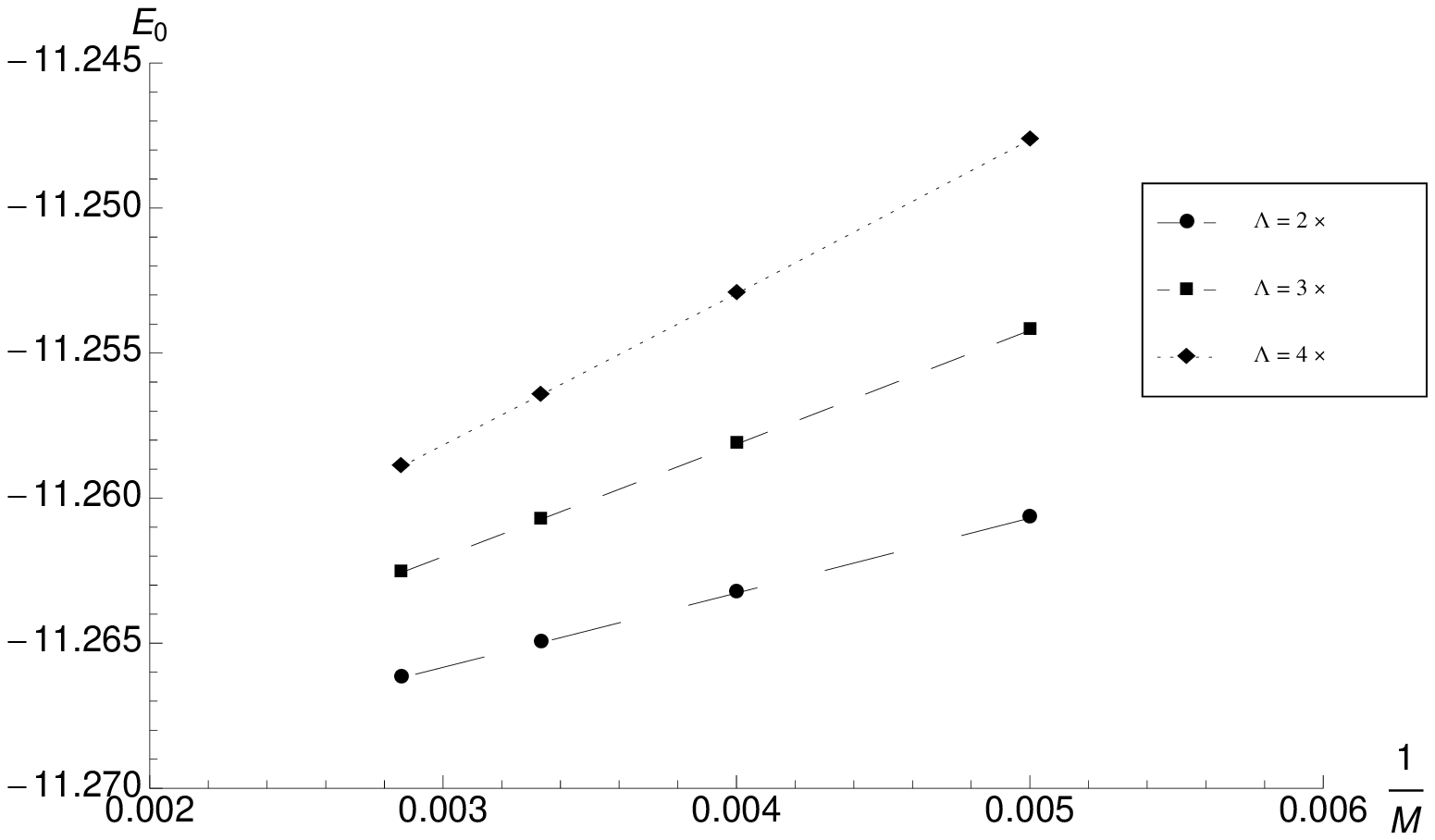}
\caption{
$L=10^{-1}$. Ground state energy, after convergence, as a function of $1/M$, the discretization roughness, for three values of the cutoff $\Lambda$. The notation 
of the legend is $\Lambda = n_{\Lambda}\times \cdots$.}
}

\newpage
\FIGURE{
\label{fig:GS-1-density}
\epsfig{width=12cm,file=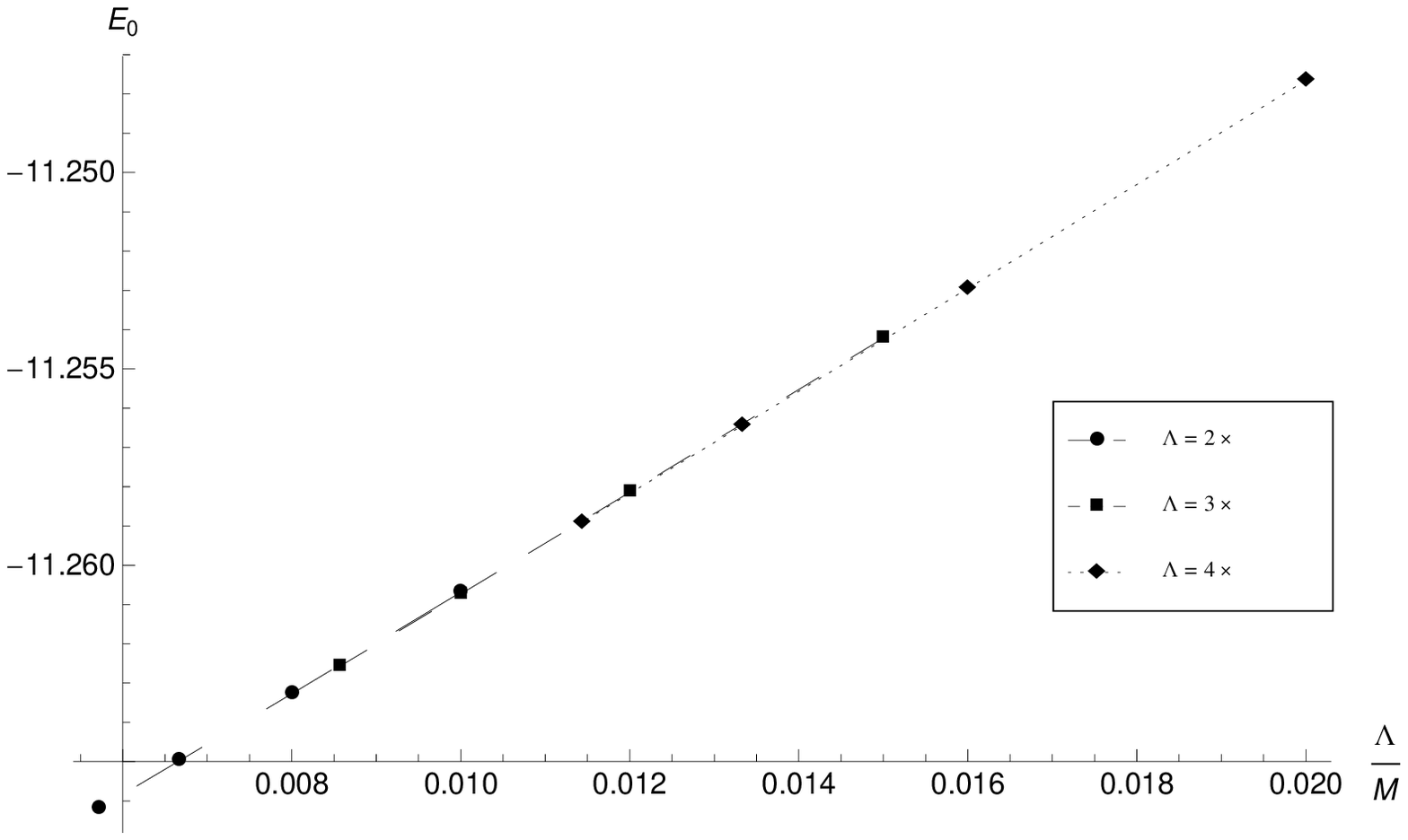}
\caption{
$L=10^{-1}$. Ground state energy, after convergence, as a function of $\Lambda/M$, the discretization density, for three values of the cutoff $\Lambda$. The notation 
of the legend is $\Lambda = n_{\Lambda}\times \cdots$. }
}

\FIGURE{
\label{fig:GS-1-profile}
\epsfig{width=14cm,file=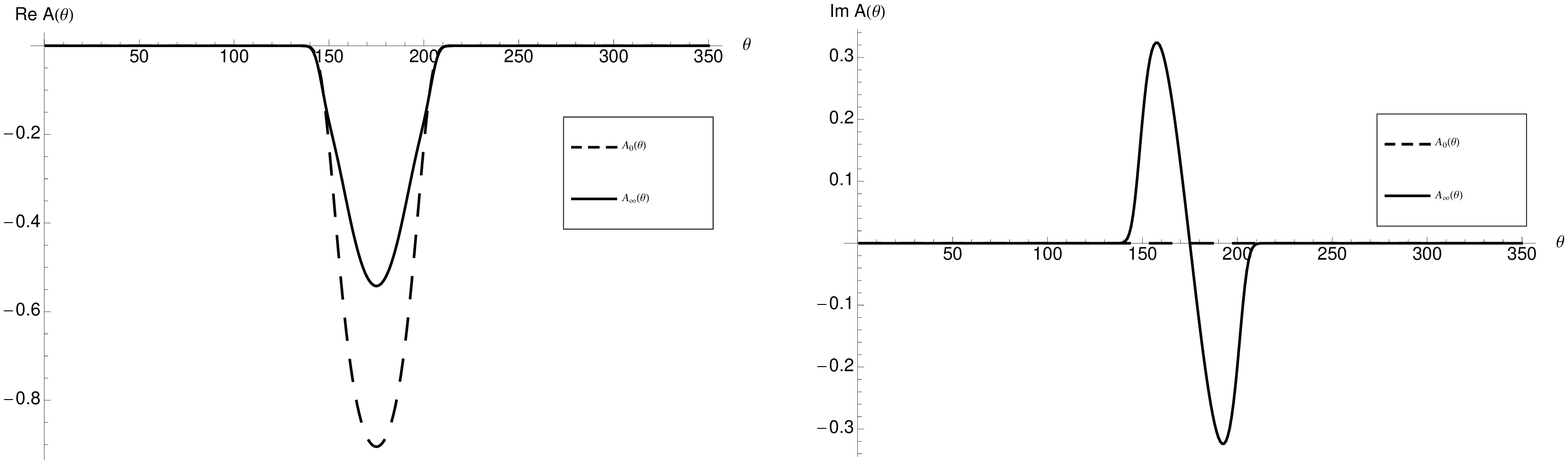}
\caption{
$L=10^{-1}$. Profile of $A(\theta)$ for the ground state, before and after convergence.
}
}

\newpage
\FIGURE{
\label{fig:GS-2-relaxation}
\epsfig{width=12cm,file=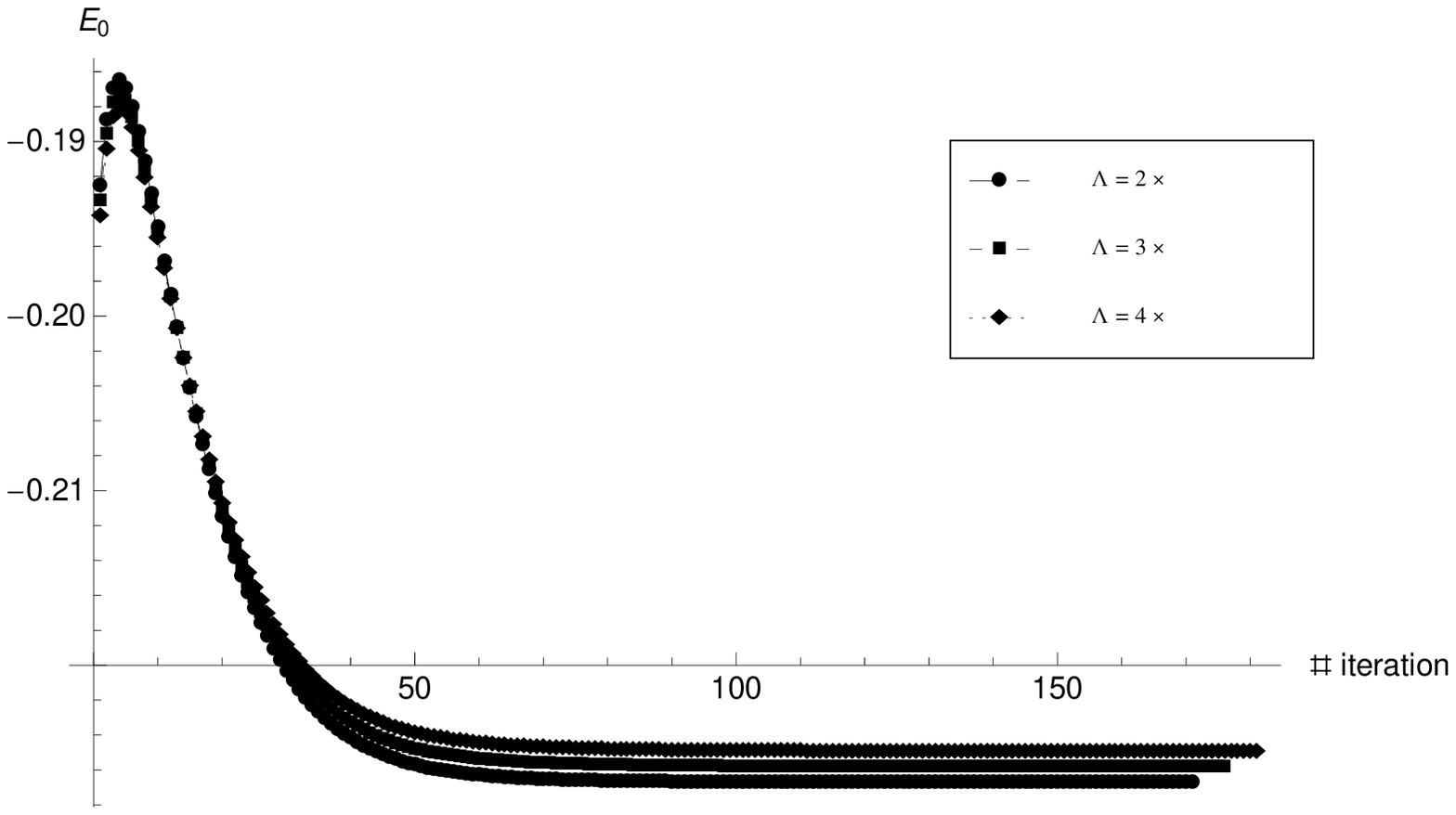}
\caption{
$L=10^{-6}$, $M=500$. Time history of the ground state energy iterates with relaxation $\lambda = 1/10$. The notation 
of the legend is $\Lambda = n_{\Lambda}\times \cdots$. The three values of $n_{\Lambda}$ are used to check
 cutoff independence.
}
}

\FIGURE{
\label{fig:GS-2-lines}
\epsfig{width=12cm,file=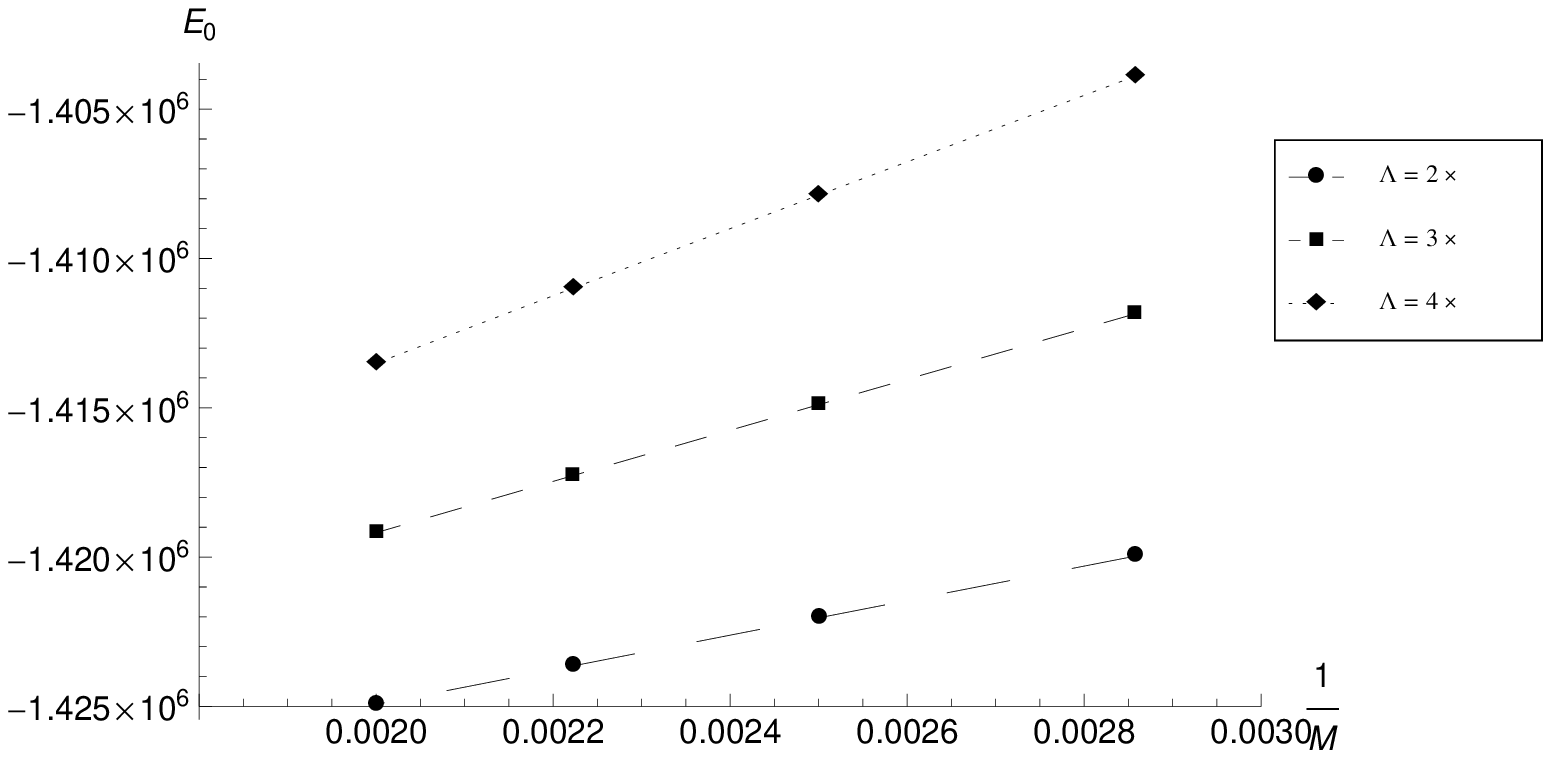}
\caption{
$L=10^{-6}$. Ground state energy, after convergence, as a function of $1/M$, the discretization roughness, for three values of the cutoff $\Lambda$. The notation 
of the legend is $\Lambda = n_{\Lambda}\times \cdots$.}
}

\newpage
\FIGURE{
\label{fig:GS-2-density}
\epsfig{width=12cm,file=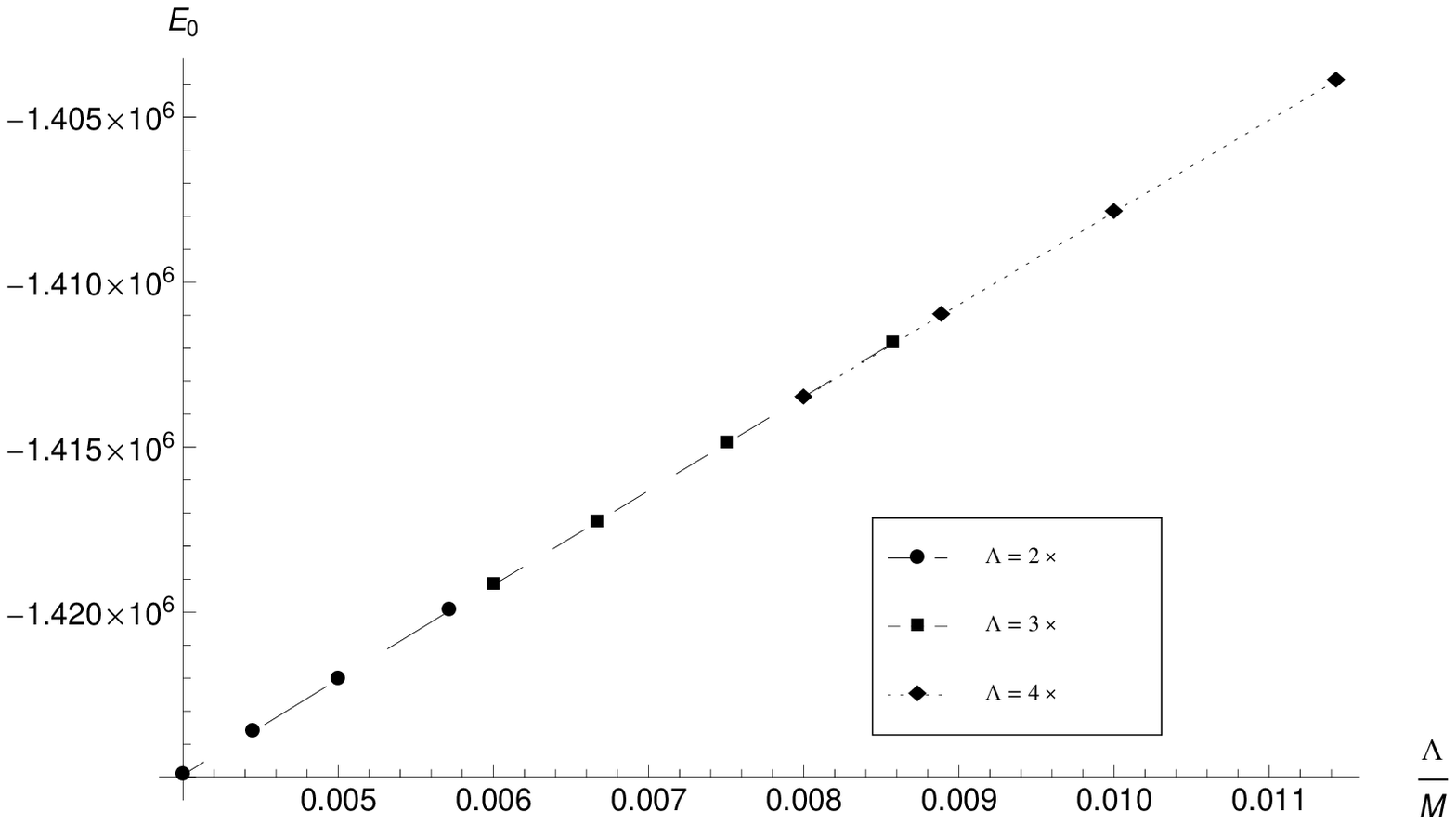}
\caption{
$L=10^{-6}$. Ground state energy, after convergence, as a function of $\Lambda/M$, the discretization density, for three values of the cutoff $\Lambda$. The notation 
of the legend is $\Lambda = n_{\Lambda}\times \cdots$. }
}

\FIGURE{
\label{fig:GS-2-profile}
\epsfig{width=14cm,file=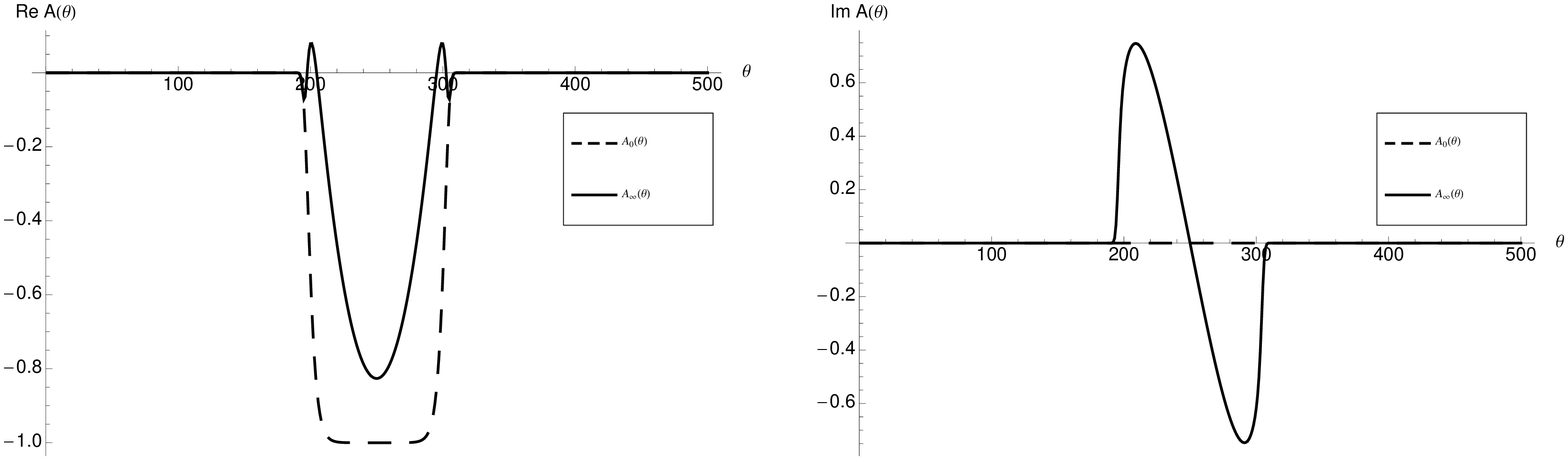}
\caption{
$L=10^{-6}$. Profile of $A(\theta)$ for the ground state, before and after convergence.
}
}

\newpage
\FIGURE{
\label{fig:GS-2-lambda}
\epsfig{width=16cm,file=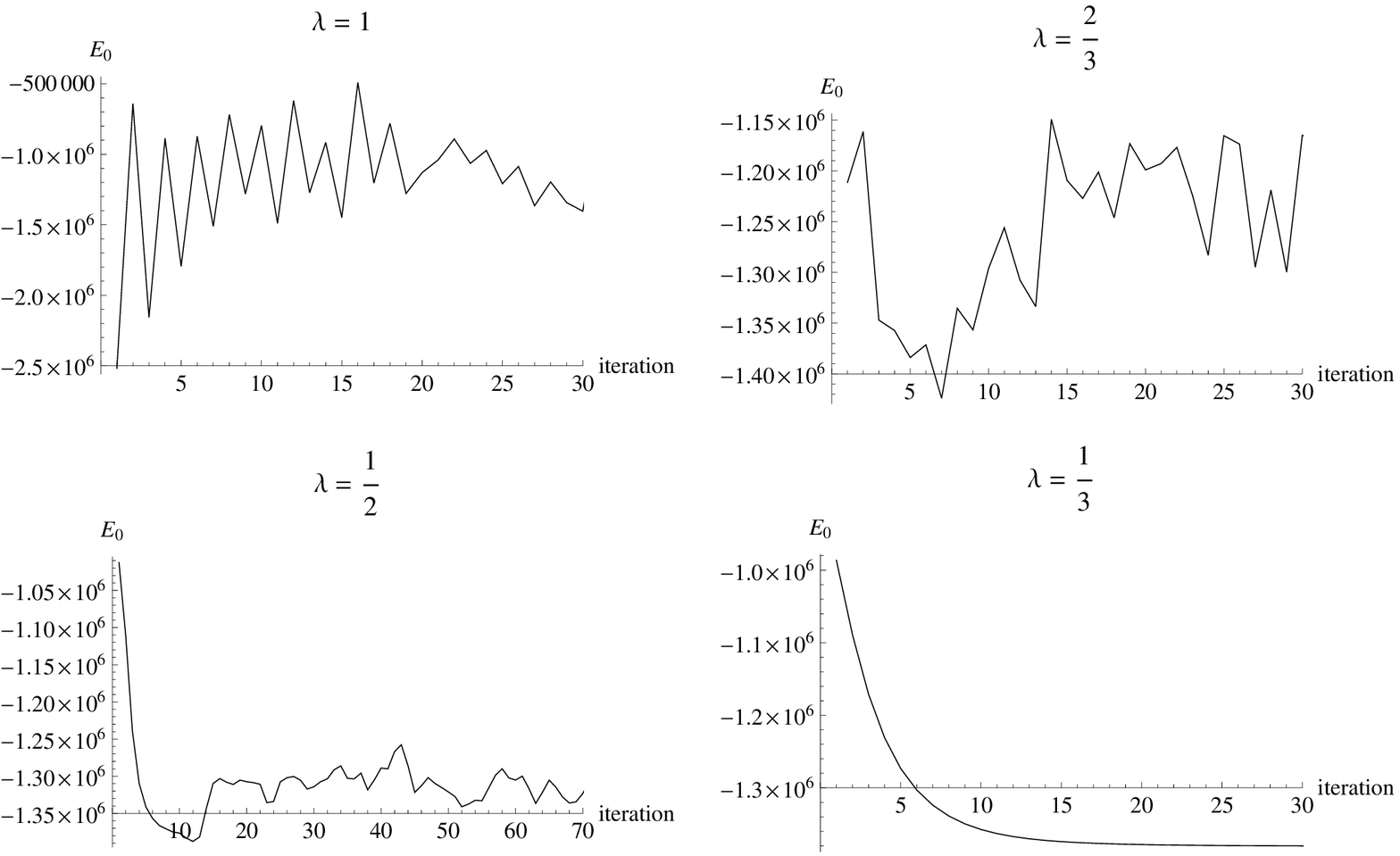}
\caption{
Convergence of the energy of the $\Theta_{0}$ state at $L=10^{-6}$, $M=100$, and $n_{\Lambda}=2$, for
various values of the relaxation parameter $\lambda$. The computation is done with $250$ digits. When the plots
start oscillating wildly, convergence is lost and the numerical accuracy rapidly decreases. The plateau which is observed in the first phase of the evolution at $\lambda=7/10$ is in agreement with the convergence at $\lambda=1/2$.
}
}

\newpage
\FIGURE{
\label{fig:Theta0-1-relaxation}
\epsfig{width=12cm,file=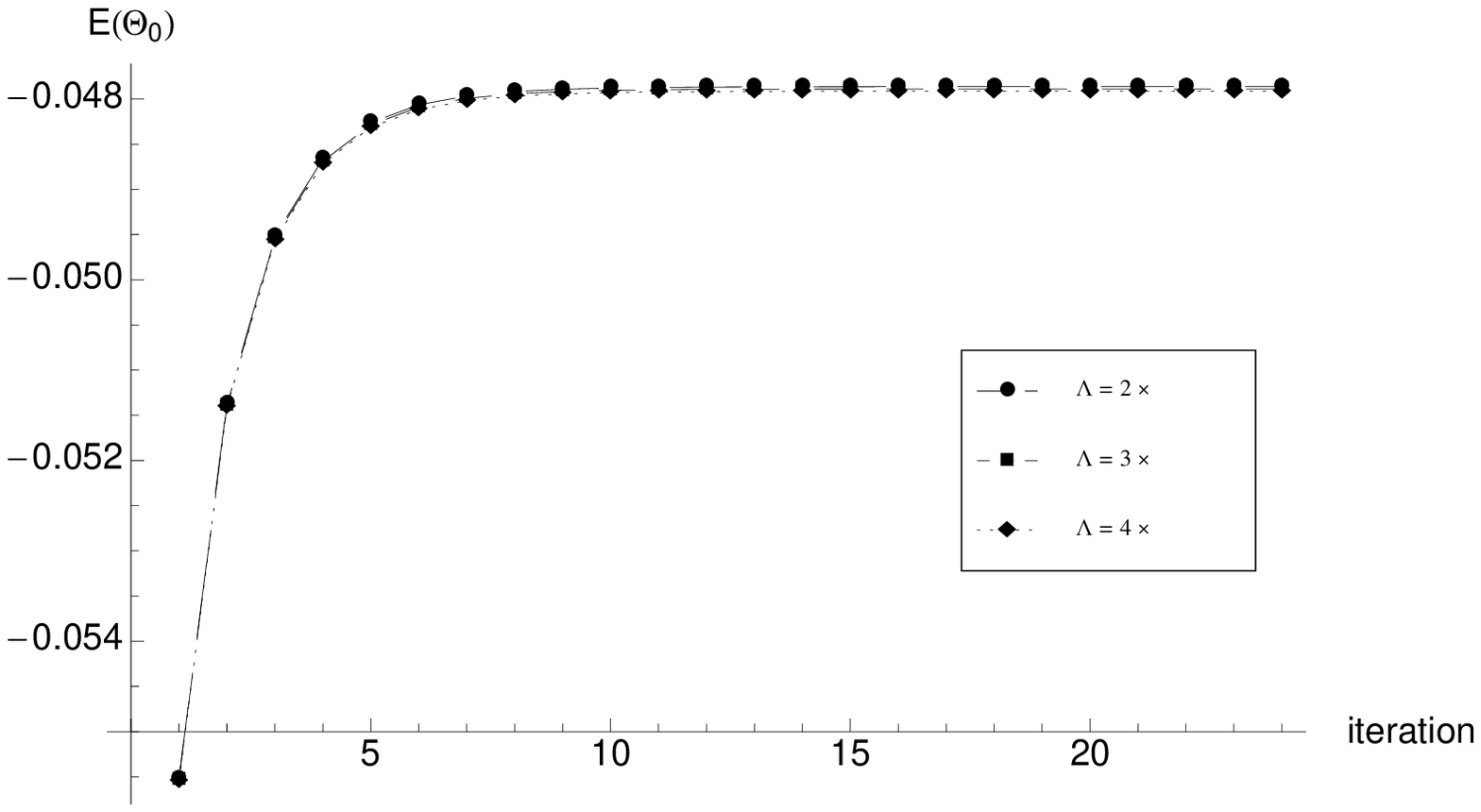}
\caption{
$L=10^{-1}$, $M=350$. Time history of the $\Theta_{0}$ state energy iterates with relaxation $\lambda = 1/2$. The notation  of the legend is $\Lambda = n_{\Lambda}\times \cdots$. The three values of $n_{\Lambda}$ are used to check
 cutoff independence.
}
}

\FIGURE{
\label{fig:Theta0-1-lines}
\epsfig{width=12cm,file=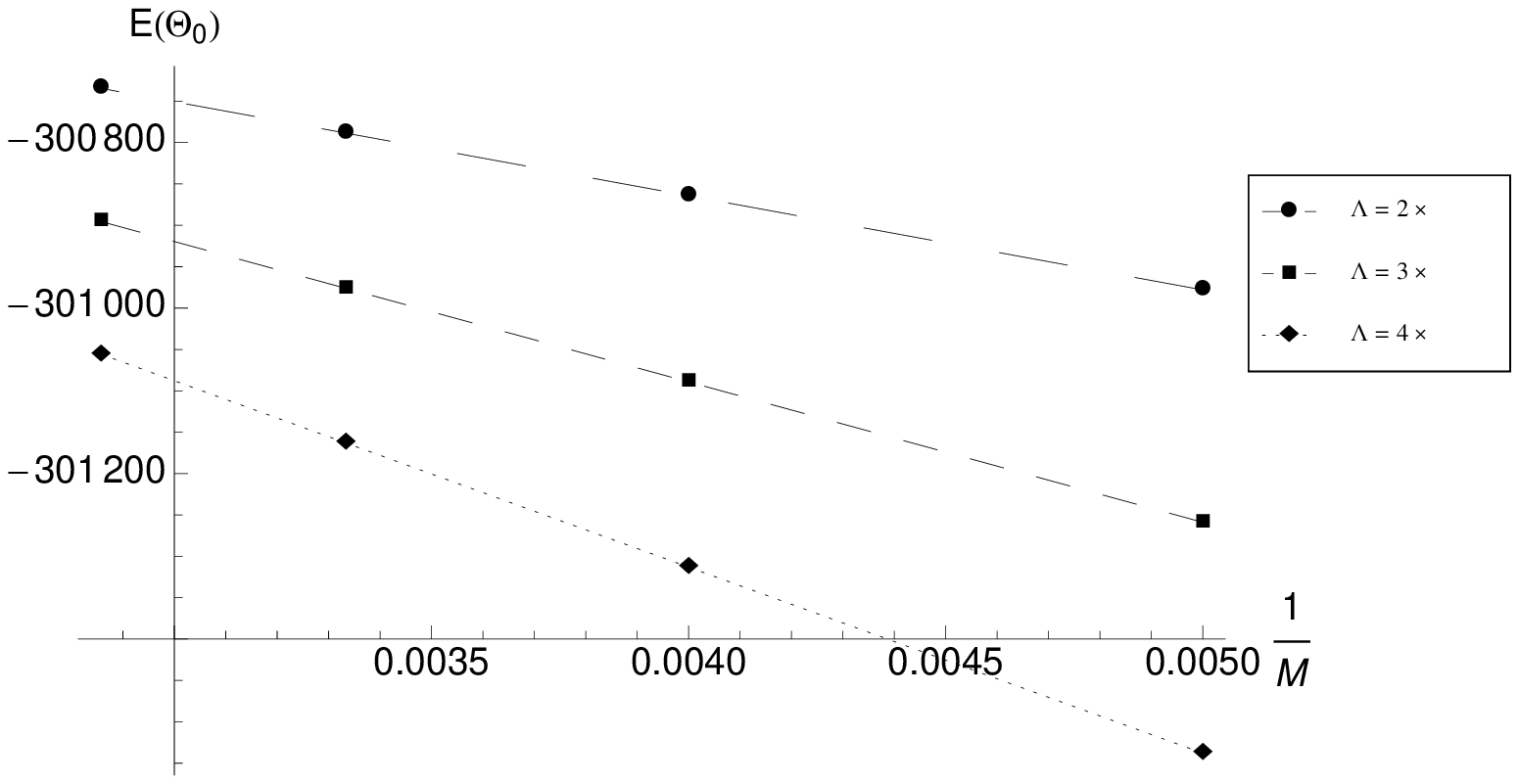}
\caption{
$L=10^{-1}$. $\Theta_{0}$ state energy, after convergence, as a function of $1/M$, the discretization roughness, for three values of the cutoff $\Lambda$. The notation 
of the legend is $\Lambda = n_{\Lambda}\times \cdots$.}
}

\newpage
\FIGURE{
\label{fig:Theta0-1-density}
\epsfig{width=12cm,file=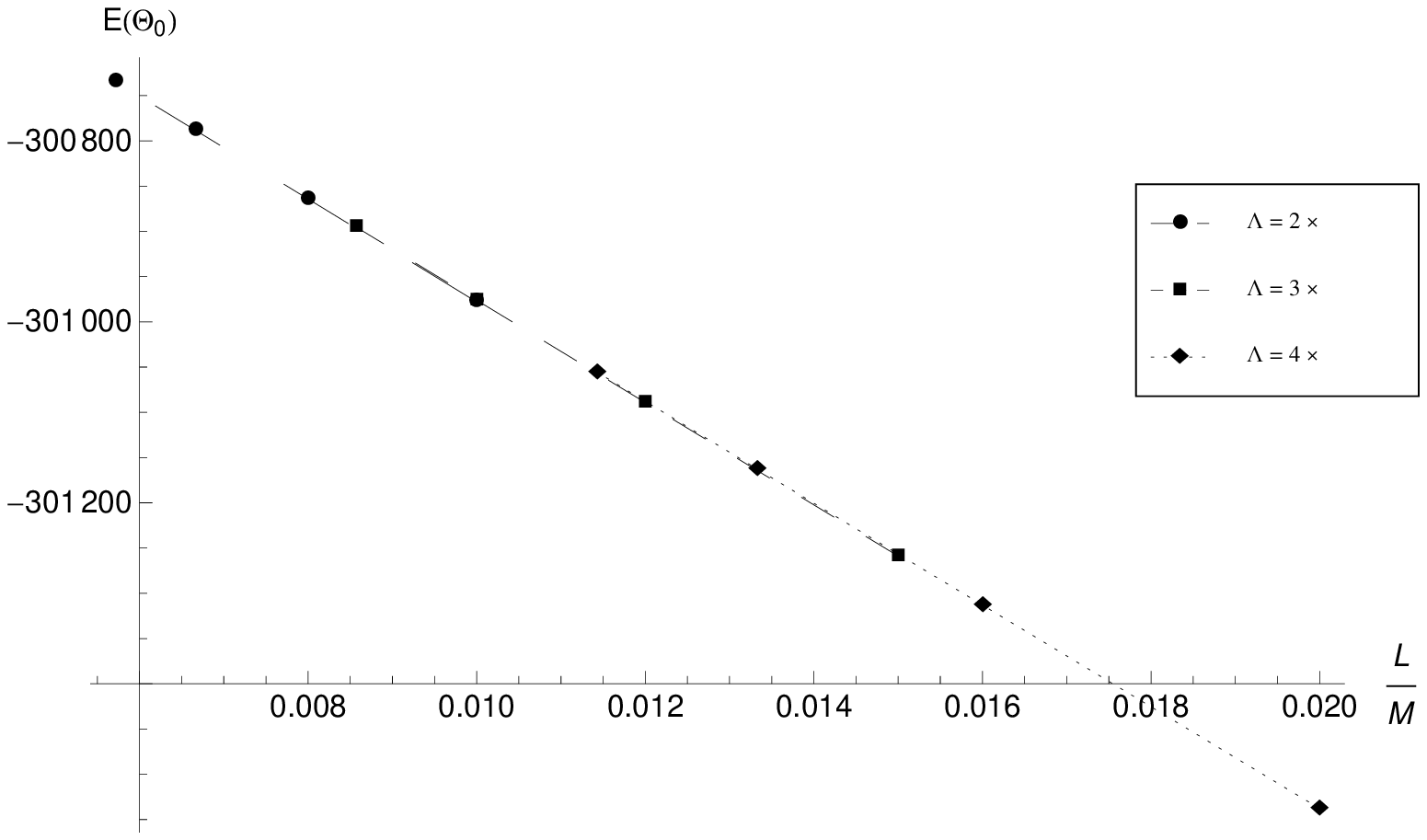}
\caption{
$L=10^{-1}$. $\Theta_{0}$ state energy, after convergence, as a function of $\Lambda/M$, the discretization density, for three values of the cutoff $\Lambda$. The notation 
of the legend is $\Lambda = n_{\Lambda}\times \cdots$. }
}

\FIGURE{
\label{fig:Theta0-2-relaxation}
\epsfig{width=12cm,file=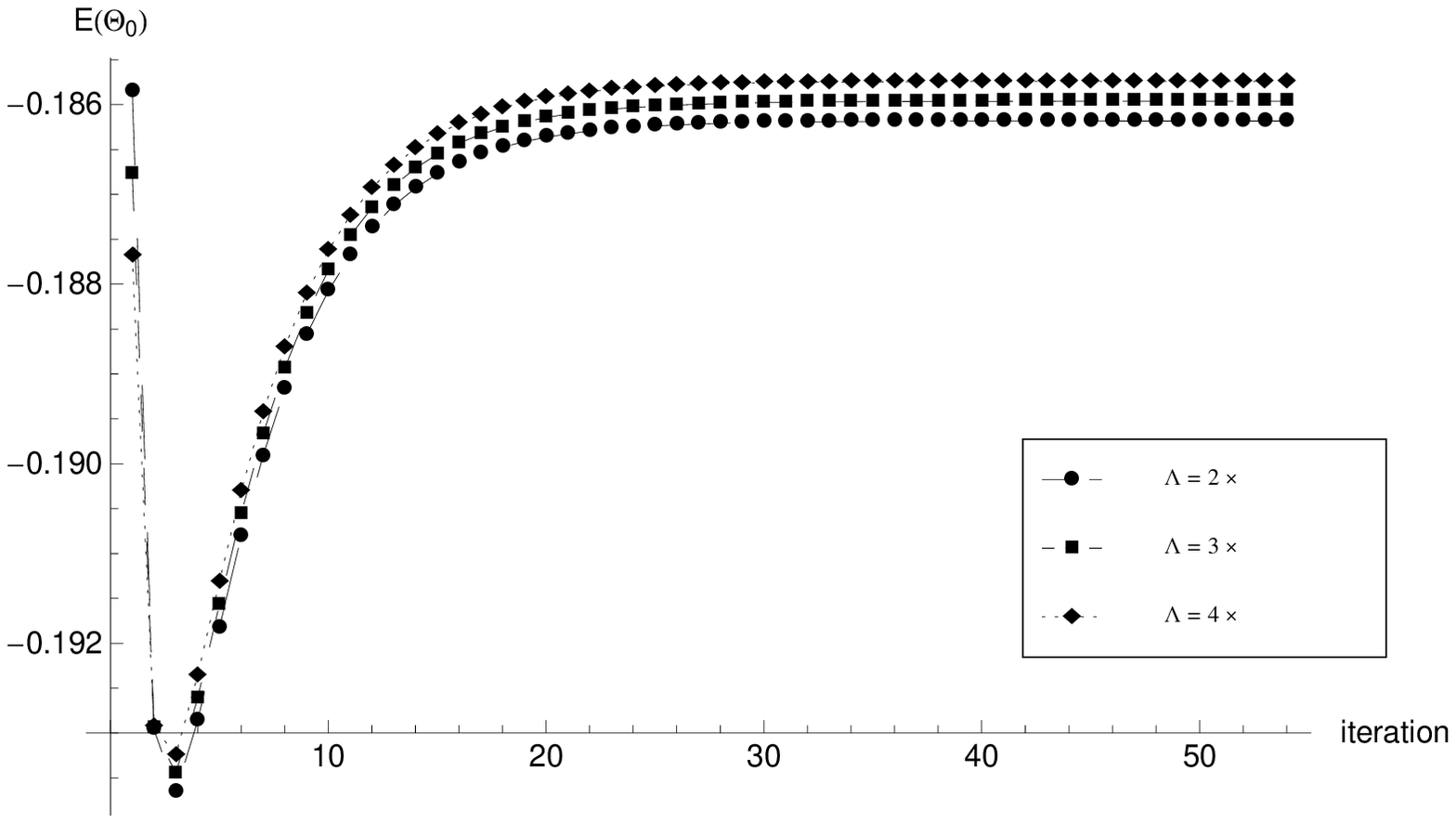}
\caption{
$L=10^{-6}$, $M=600$. Time history of the $\Theta_{0}$ state energy iterates with relaxation $\lambda = 1/2$. The notation  of the legend is $\Lambda = n_{\Lambda}\times \cdots$. The three values of $n_{\Lambda}$ are used to check
 cutoff independence.
}
}

\FIGURE{
\label{fig:Theta0-2-lines}
\epsfig{width=12cm,file=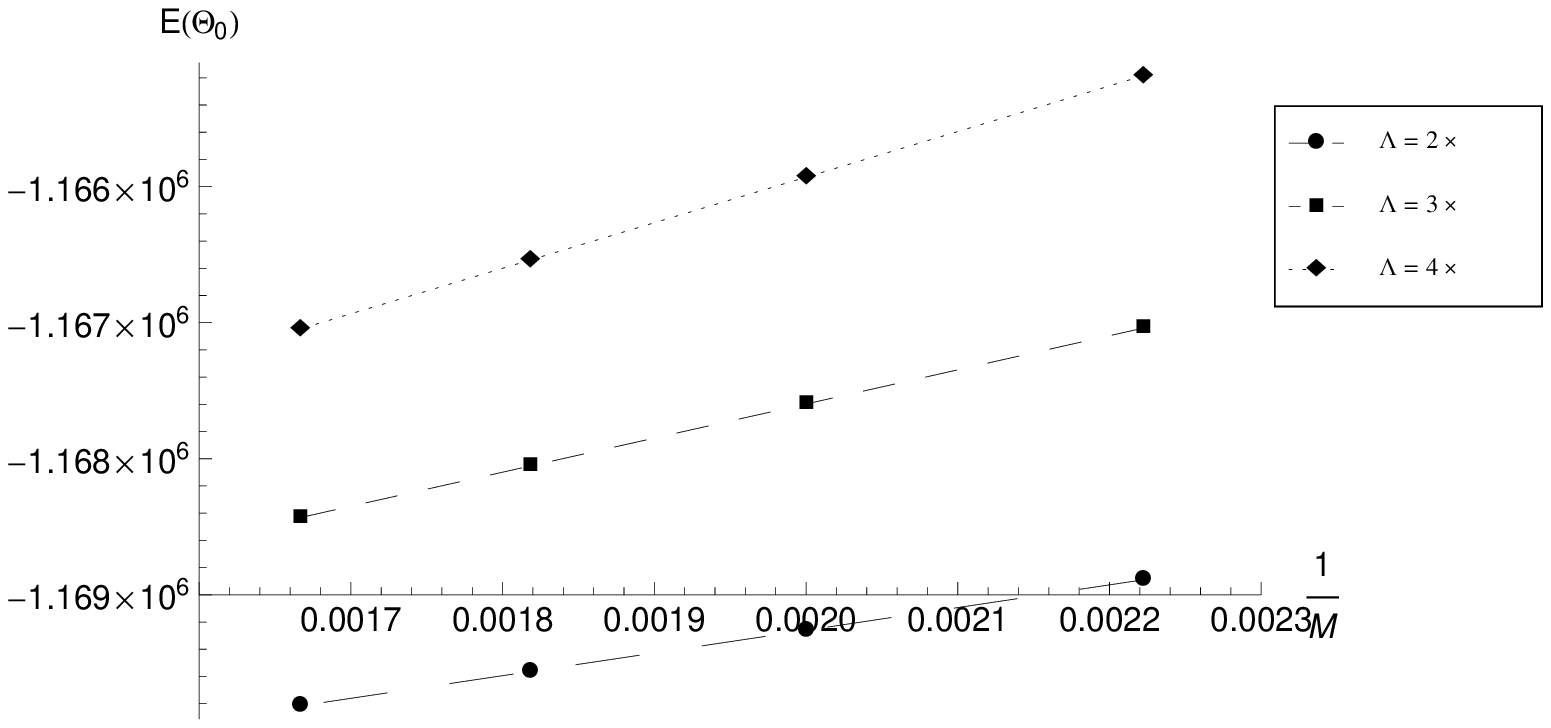}
\caption{
$L=10^{-6}$. $\Theta_{0}$ state energy, after convergence, as a function of $1/M$, the discretization roughness, for three values of the cutoff $\Lambda$. The notation 
of the legend is $\Lambda = n_{\Lambda}\times \cdots$.}
}

\FIGURE{
\label{fig:Theta0-2-density}
\epsfig{width=12cm,file=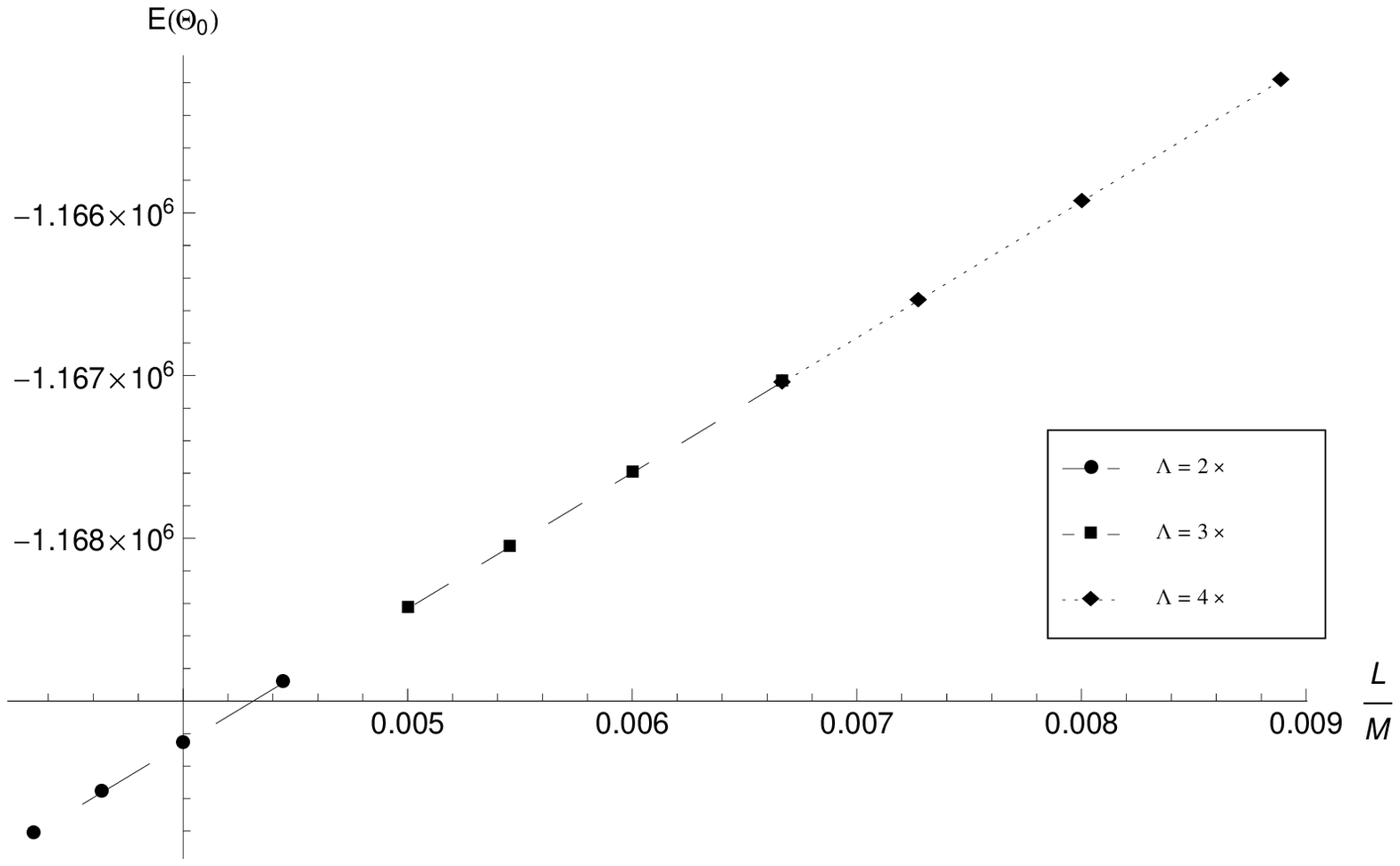}
\caption{
$L=10^{-6}$. $\Theta_{0}$ state energy, after convergence, as a function of $\Lambda/M$, the discretization density, for three values of the cutoff $\Lambda$. The notation 
of the legend is $\Lambda = n_{\Lambda}\times \cdots$. }
}

\FIGURE{
\label{fig:Theta0-2-profile}
\epsfig{width=14cm,file=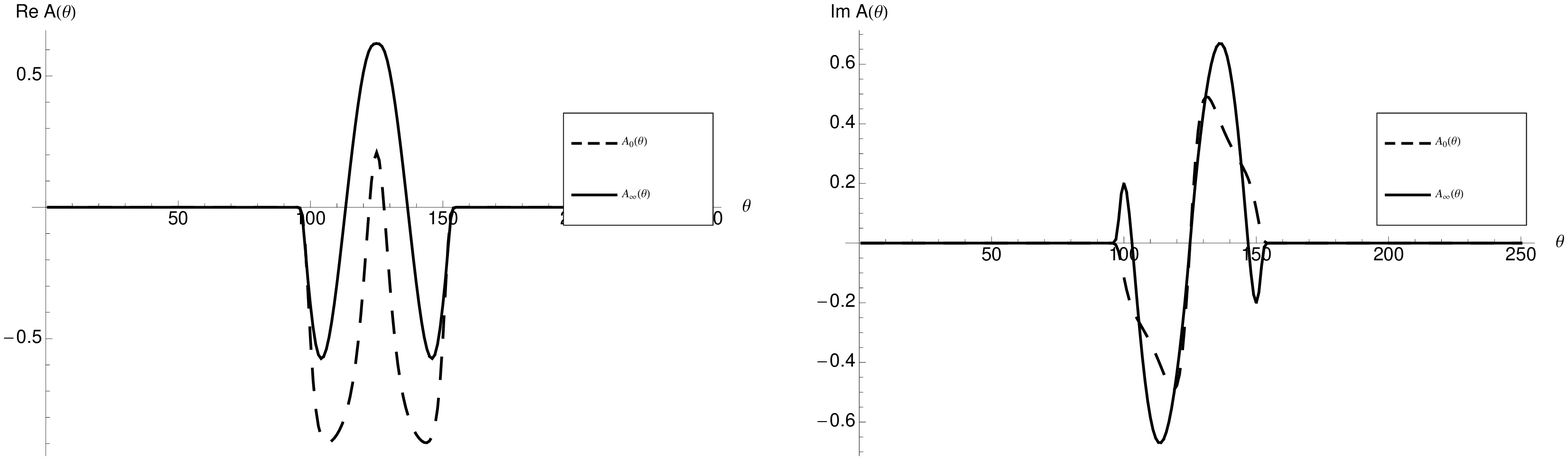}
\caption{
$L=10^{-6}$. Profile of $A(\theta)$ for the $\Theta_{0}$ state, before and after convergence.
}
}

\FIGURE{
\label{fig:Theta0-2-lambda}
\epsfig{width=16cm,file=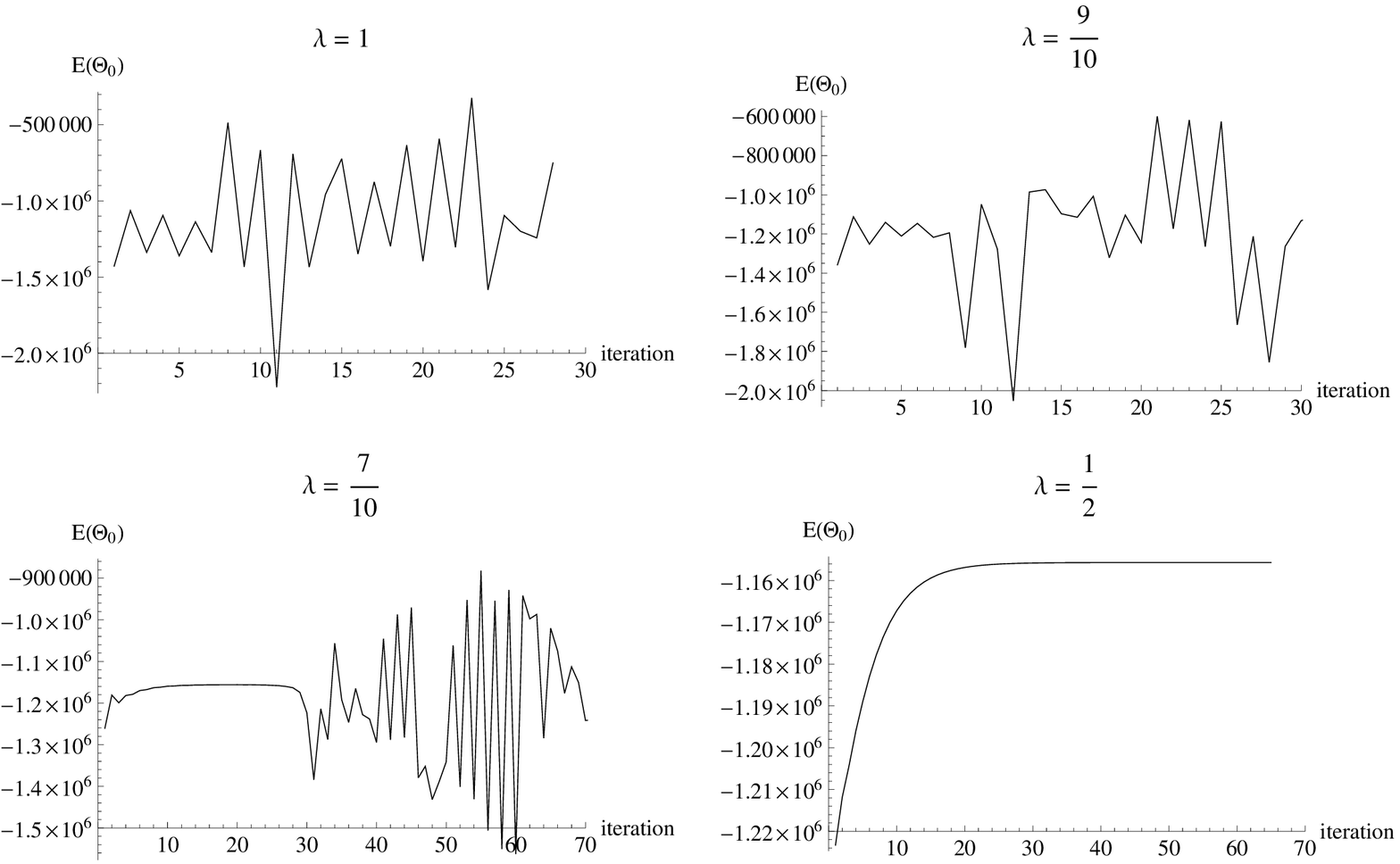}
\caption{
Convergence of the energy of the $\Theta_{0}$ state at $L=10^{-6}$, $M=100$, and $n_{\Lambda}=2$, for
various values of the relaxation parameter $\lambda$. The computation is done with $250$ digits. When the plots
start oscillating wildly, convergence is lost and the numerical accuracy rapidly decreases. The plateau which is observed in the first phase of the evolution at $\lambda=7/10$ is in agreement with the convergence at $\lambda=1/2$.
}
}

\FIGURE{
\label{fig:Theta00-2-energy-lambda}
\epsfig{width=14cm,file=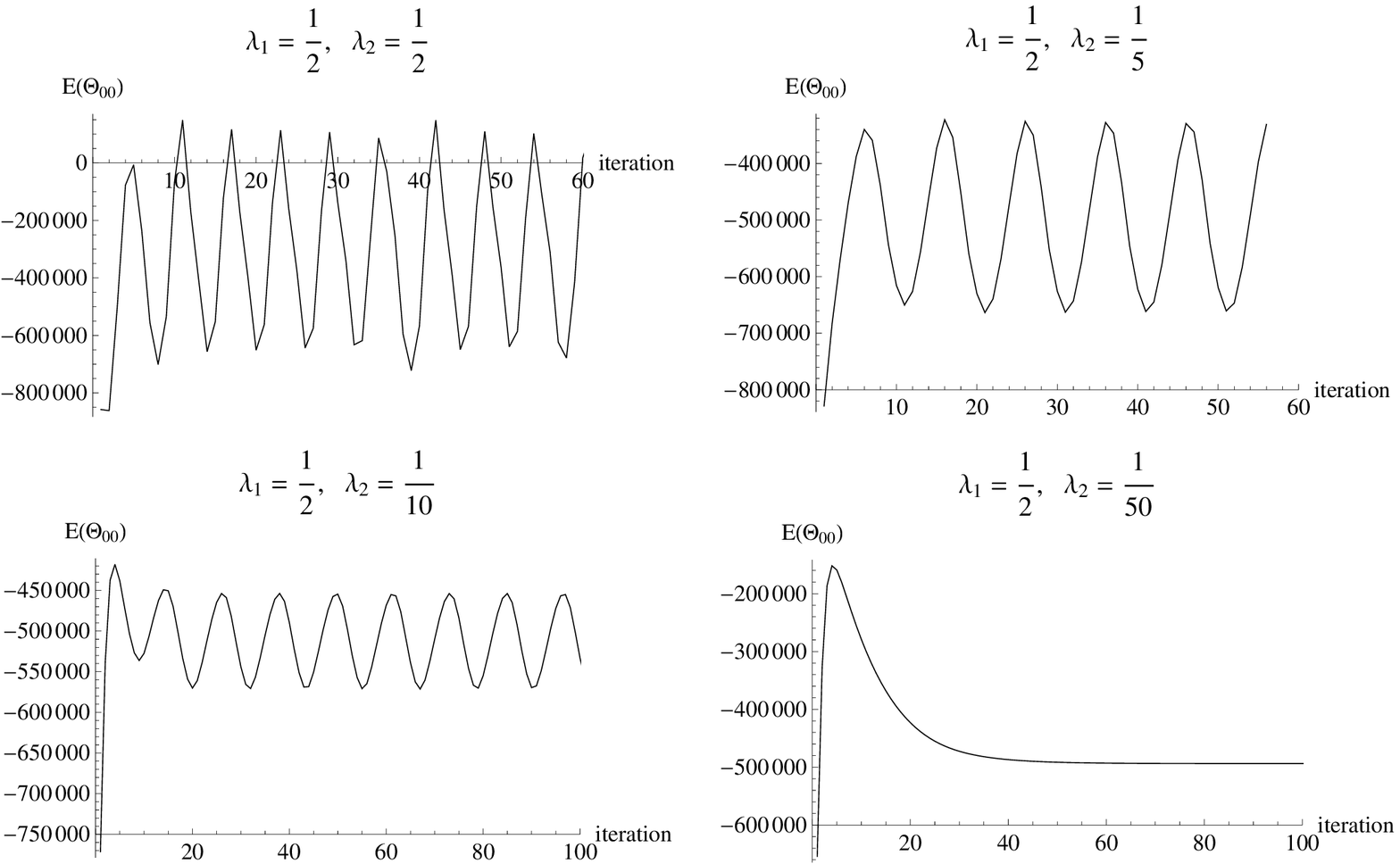}
\caption{
Convergence of the energy of the $\Theta_{00}$ state at $L=10^{-6}$, $M=50$, and $n_{\Lambda}=2$, for
$\lambda_{1}=\frac{1}{2}$ and 
various values of the relaxation parameter $\lambda_{2}$. 
}
}

\FIGURE{
\label{fig:Theta00-2-theta-lambda}
\epsfig{width=14cm,file=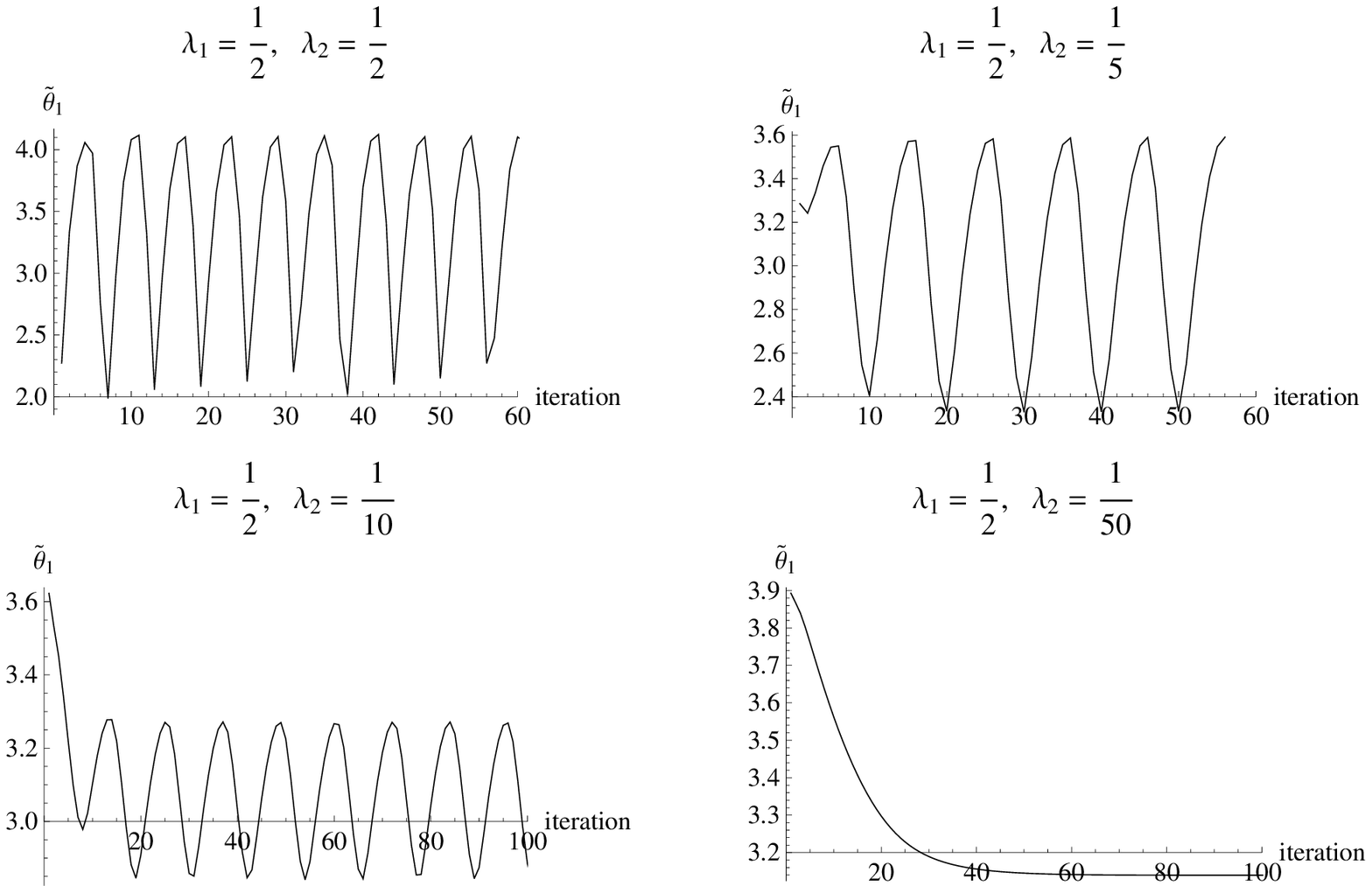}
\caption{
Convergence of the Bethe root $\widetilde\theta_{1}$ at $L=10^{-6}$, $M=50$, and $n_{\Lambda}=2$, for
$\lambda_{1}=\frac{1}{2}$ and 
various values of the relaxation parameter $\lambda_{2}$. 
}
}

\FIGURE{
\label{fig:Theta00-2-lines}
\epsfig{width=12cm,file=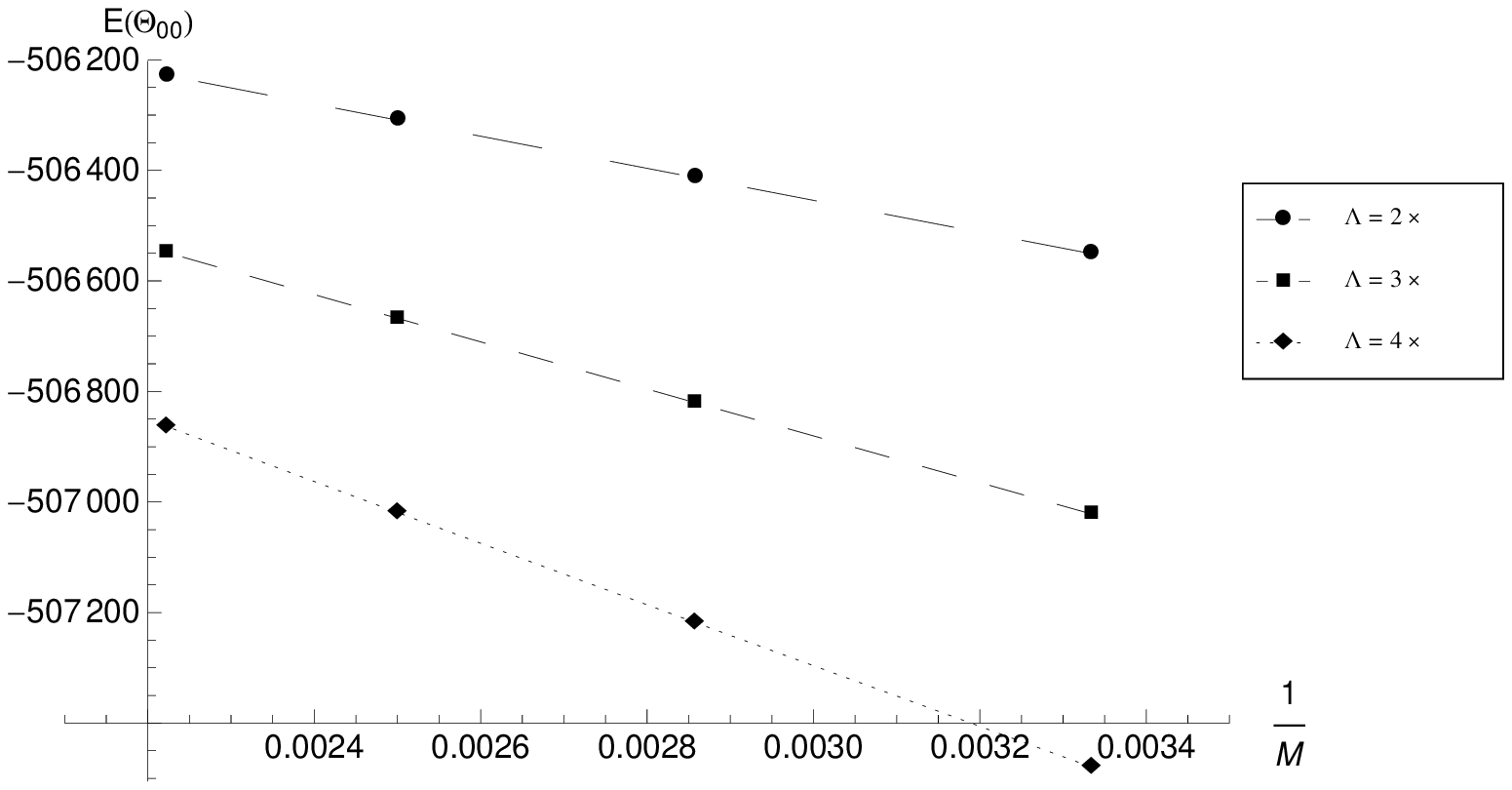}
\caption{
$L=10^{-6}$. $\Theta_{00}$ state energy, after convergence, as a function of $1/M$, the discretization roughness, for three values of the cutoff $\Lambda$. The notation 
of the legend is $\Lambda = n_{\Lambda}\times \cdots$.}
}

\FIGURE{
\label{fig:Theta00-2-density}
\epsfig{width=12cm,file=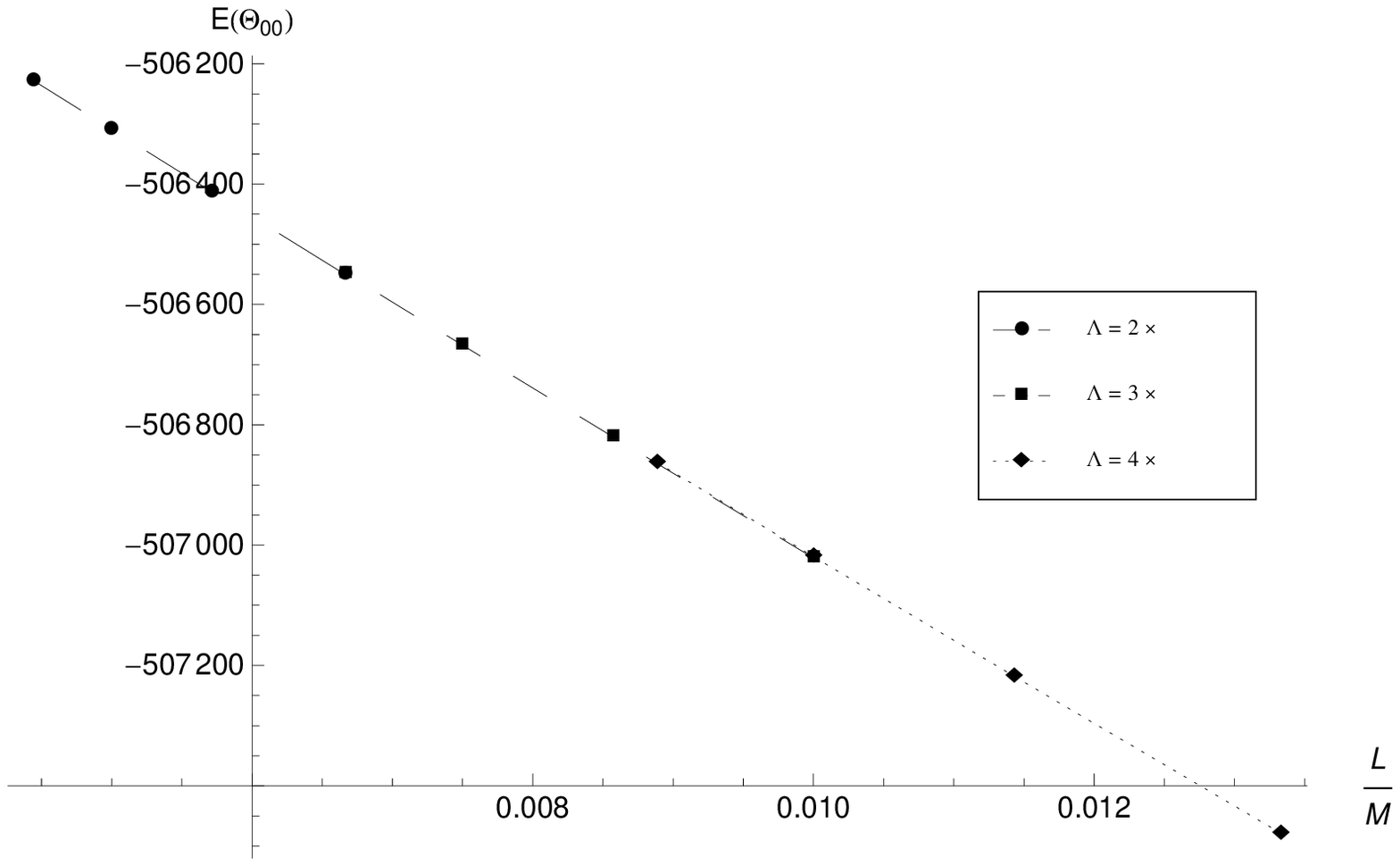}
\caption{
$L=10^{-6}$. $\Theta_{00}$ state energy, after convergence, as a function of $\Lambda/M$, the discretization density, for three values of the cutoff $\Lambda$. The notation 
of the legend is $\Lambda = n_{\Lambda}\times \cdots$. }
}

\FIGURE{
\label{fig:Theta00-2-profile}
\epsfig{width=14cm,file=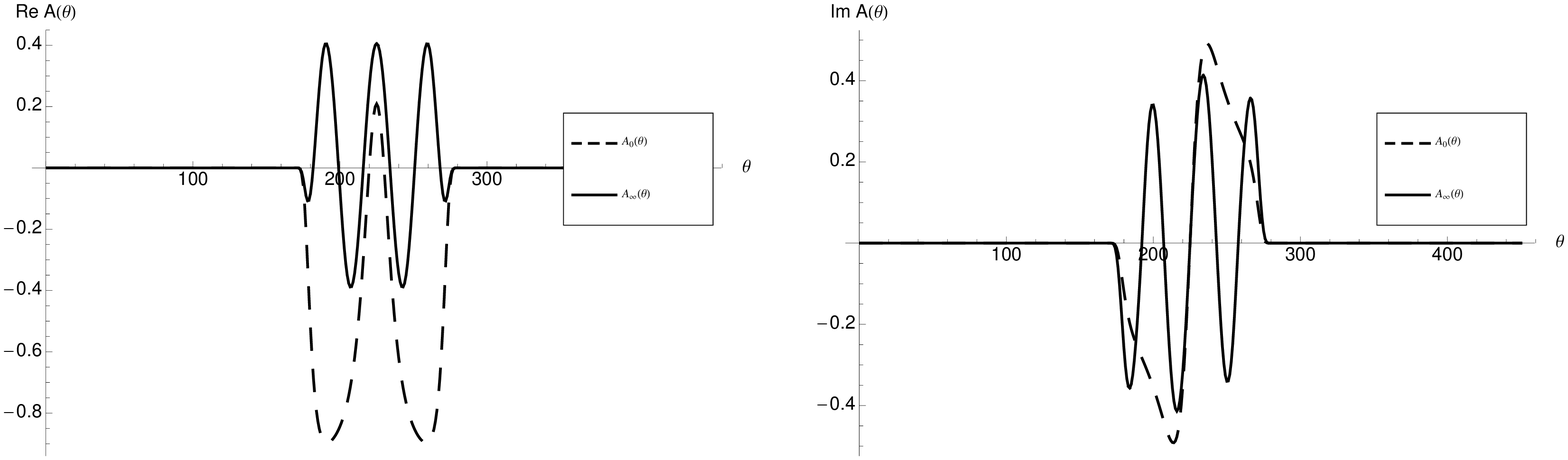}
\caption{
$L=10^{-6}$. Profile of $A(\theta)$ for the $\Theta_{00}$ state, before and after convergence.
}
}

\FIGURE{
\label{fig:Theta00-all-profile}
\epsfig{width=16cm,file=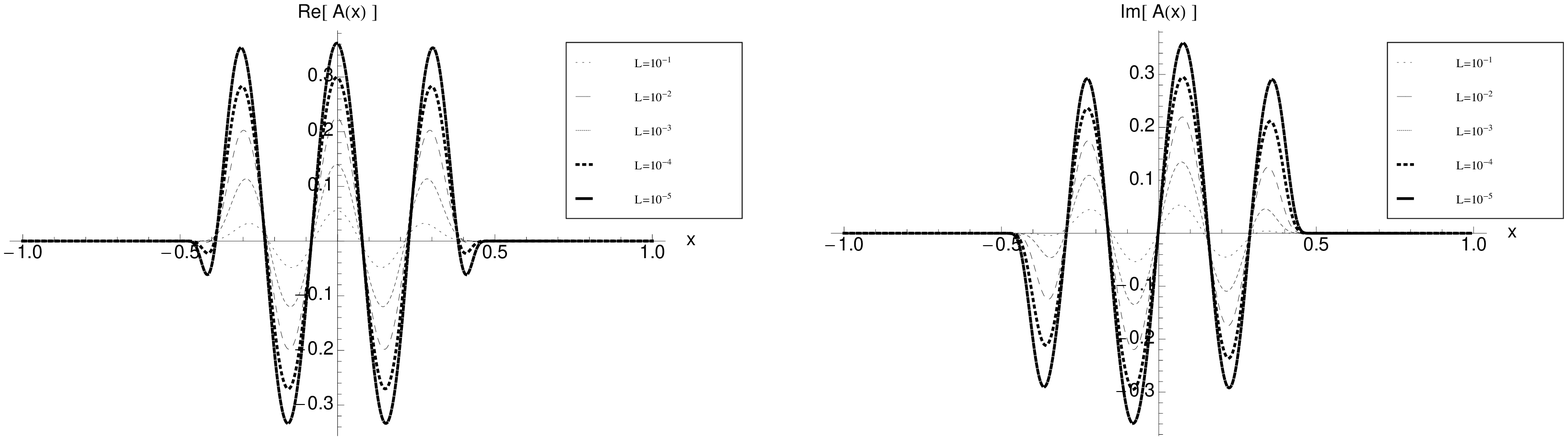}
\caption{
Profile of $A(\theta)$ for the $\Theta_{00}$ state after convergence at various sizes $L$.
}
}

\FIGURE{
\label{fig:final-plot}
\epsfig{width=16cm,file=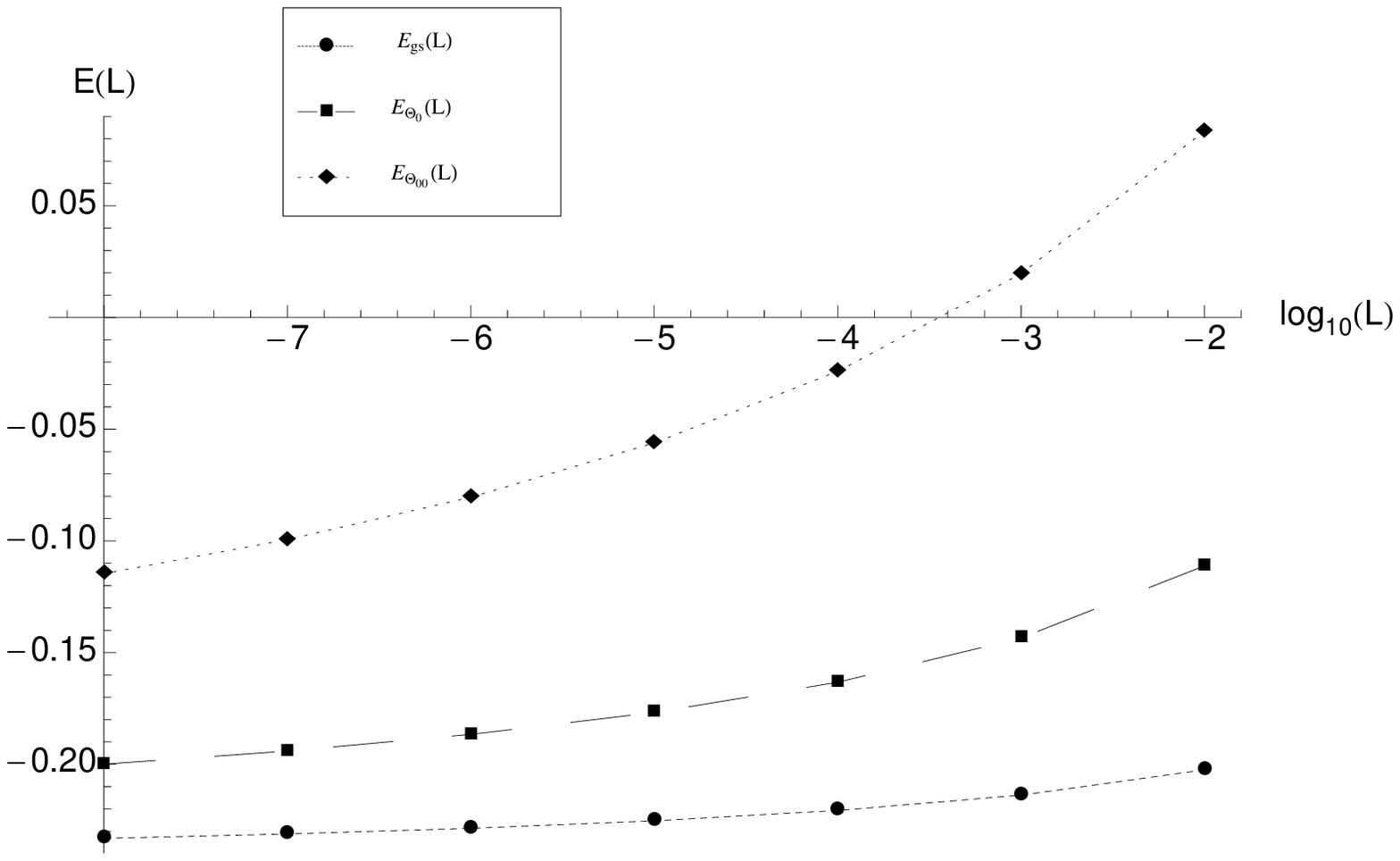}
\caption{
Summary plot showing the size dependence of the energies of the three considered states.
}
}


\begin{thebibliography}{99}

\bibitem{Duality}
  J.~M.~Maldacena,
  {\em The Large N limit of superconformal field theories and supergravity},
  Adv.\ Theor.\ Math.\ Phys.\  {\bf 2}, 231-252 (1998).
  [hep-th/9711200].


\bibitem{JPA}
See the special issue {\em Integrability and the AdS/CFT correspondence},
Guest Editors: A. A. Tseytlin, C. Kristjansen, and M. Staudacher, J.\ Phys.\ A {\bf A42}, 254002 (2009).

\bibitem{Wrapping}
  C.~Sieg, A.~Torrielli,
  {\em Wrapping interactions and the genus expansion of the 2-point function of composite operators},
  Nucl.\ Phys.\  {\bf B723}, 3-32 (2005).
  [hep-th/0505071].
  
  

\bibitem{Luscher}
  M.~Luscher,
  {\em Volume Dependence of the Energy Spectrum in Massive Quantum Field Theories. 1. Stable Particle States},
  Commun.\ Math.\ Phys.\  {\bf 104}, 177 (1986) \blt
  M.~Luscher,
  {\em Volume Dependence of the Energy Spectrum in Massive Quantum Field Theories. 2. Scattering States},
  Commun.\ Math.\ Phys.\  {\bf 105}, 153-188 (1986).



\bibitem{ddv}
  C.~Destri, H.~J.~de Vega,
  {\em Light Cone Lattices And The Exact Solution Of Chiral Fermion And Sigma Models},
  J.\ Phys.\ A {\bf A22}, 1329 (1989) \blt
  C.~Destri, H.~J.~de Vega,
  {\em Integrable Quantum Field Theories And Conformal Field Theories From Lattice Models In The Light Cone Approach},
  Phys.\ Lett.\  {\bf B201}, 261 (1988) \blt
  C.~Destri, H.~J.~de Vega,
  {\em Light Cone Lattice Approach To Fermionic Theories In 2-d: The Massive Thirring Model},
  Nucl.\ Phys.\  {\bf B290}, 363 (1987).




\bibitem{TBA}
  C.~Destri, H.~J.~de Vega,
 {\em Light Cone Lattices And The Exact Solution Of Chiral Fermion And Sigma Models},
  J.\ Phys.\ A {\bf A22}, 1329 (1989) \blt 
 %
  A.~B.~Zamolodchikov,
  {\em Thermodynamic Bethe Ansatz In Relativistic Models. Scaling Three State Potts And Lee-Yang Models},
  Nucl.\ Phys.\  {\bf B342}, 695-720 (1990)  \blt
%
  V.~V.~Bazhanov, S.~L.~Lukyanov, A.~B.~Zamolodchikov,
  {\em Integrable quantum field theories in finite volume: Excited state energies},
  Nucl.\ Phys.\  {\bf B489}, 487-531 (1997)
  [hep-th/9607099] \blt
%
  P.~Dorey, R.~Tateo,
  {\em Excited states by analytic continuation of TBA equations},
  Nucl.\ Phys.\  {\bf B482}, 639-659 (1996)
  [hep-th/9607167] \blt
%
%
  P.~Dorey, R.~Tateo,
  {\em Anharmonic oscillators, the thermodynamic Bethe ansatz, and nonlinear integral equations},
  J.\ Phys.\ A {\bf A32}, L419-L425 (1999) \blt
  [hep-th/9812211].
  D.~Fioravanti, A.~Mariottini, E.~Quattrini, and F. ~Ravanini,
  {\em Excited state Destri-De Vega equation for Sine-Gordon and restricted Sine-Gordon models},
  Phys.\ Lett.\  {\bf B390}, 243-251 (1997)
  [hep-th/9608091] \blt
  %
  V.~V.~Bazhanov, S.~L.~Lukyanov, A.~B.~Zamolodchikov,
  {\em Integrable structure of conformal field theory, quantum KdV theory and thermodynamic Bethe Ansatz},
  Commun.\ Math.\ Phys.\  {\bf 177}, 381-398 (1996)
  [hep-th/9412229] \blt
%
  V.~V.~Bazhanov, S.~L.~Lukyanov, A.~B.~Zamolodchikov,
  {\em Integrable structure of conformal field theory. 3. The Yang-Baxter relation}
  Commun.\ Math.\ Phys.\  {\bf 200}, 297-324 (1999)
  [hep-th/9805008] \blt
%
  J.~Teschner,
  {\em On the spectrum of the Sinh-Gordon model in finite volume},
  Nucl.\ Phys.\  {\bf B799}, 403-429 (2008)
  [hep-th/0702214].

\bibitem{Hegedus:2004xd}
  A.~Hegedus,
  {\em Nonlinear integral equations for finite volume excited state energies of the O(3) and O(4) nonlinear sigma-models},
  J.\ Phys.\ A {\bf A38}, 5345-5358 (2005).
  [hep-th/0412125].

\bibitem{Balog:2003yr}
  J.~Balog, A.~Hegedus,
  {\em TBA Equations for excited states in the O(3) and O(4) nonlinear sigma model},
  J.\ Phys.\ A {\bf A37}, 1881-1901 (2004).
  [hep-th/0309009].

\bibitem{Ysystem}
  A.~B.~Zamolodchikov,
  {\em TBA equations for integrable perturbed $SU(2)_{k}\times SU(2)_{1}/ SU(2)_{k+1}$ coset models},
  Nucl.\ Phys.\  {\bf B366}, 122-134 (1991) \blt  
  A.~Kuniba, T.~Nakanishi, J.~Suzuki,
  {\em Functional relations in solvable lattice models. 1: Functional relations and representation theory},
  Int.\ J.\ Mod.\ Phys.\  {\bf A9}, 5215-5266 (1994).
  [hep-th/9309137].



\bibitem{AdS-Y}
  N.~Gromov, V.~Kazakov, P.~Vieira,
  {\em Exact Spectrum of Anomalous Dimensions of Planar N=4 Supersymmetric Yang-Mills Theory},
  Phys.\ Rev.\ Lett.\  {\bf 103}, 131601 (2009).
  [arXiv:0901.3753 [hep-th]] \blt
  D.~Bombardelli, D.~Fioravanti, R.~Tateo,
  {\em Thermodynamic Bethe Ansatz for planar AdS/CFT: A Proposal},
  J.\ Phys.\ A {\bf A42}, 375401 (2009)
  [arXiv:0902.3930 [hep-th]] \blt
%
  N.~Gromov, V.~Kazakov, A.~Kozak {\it et al.},
 {\em Exact Spectrum of Anomalous Dimensions of Planar N = 4 Supersymmetric Yang-Mills Theory: TBA and excited states},
  Lett.\ Math.\ Phys.\  {\bf 91}, 265-287 (2010)
  [arXiv:0902.4458 [hep-th]] \blt
%
  G.~Arutyunov, S.~Frolov,
  {\em Thermodynamic Bethe Ansatz for the $\ads$ Mirror Model},
  JHEP {\bf 0905}, 068 (2009)
  [arXiv:0903.0141 [hep-th]] \blt
  N.~Gromov, V.~Kazakov, P.~Vieira,
  {\em Exact AdS/CFT spectrum: Konishi dimension at any coupling},
  Phys. Rev. Lett. 104 (2010) 211601 [arXiv:0906.4240 [hep-th]].

\bibitem{K1}
  N.~Gromov, V.~Kazakov, P.~Vieira,
  {\em Finite Volume Spectrum of 2D Field Theories from Hirota Dynamics},
  JHEP {\bf 0912}, 060 (2009).
  [arXiv:0812.5091 [hep-th]].

\bibitem{K2}
  V.~Kazakov, S.~Leurent,
  {\em Finite Size Spectrum of SU(N) Principal Chiral Field from Discrete Hirota Dynamics},
    [arXiv:1007.1770 [hep-th]].

\bibitem{Balog:2005yz}
  J.~Balog, A.~Hegedus,
  {\em TBA equations for the mass gap in the O(2r) non-linear sigma-models},
  Nucl.\ Phys.\  {\bf B725}, 531-553 (2005).
  [hep-th/0504186].
  
\bibitem{Gromov:2006dh}
  N.~Gromov, V.~Kazakov, K.~Sakai {\it et al.},
  {\em Strings as multi-particle states of quantum sigma-models},
  Nucl.\ Phys.\  {\bf B764}, 15-61 (2007).
  [hep-th/0603043].

\bibitem{Zamolodchikov:1978xm}
  A.~B.~Zamolodchikov, A.~B.~Zamolodchikov,
  {\em Factorized S-Matrices in Two-Dimensions as the Exact Solutions of Certain Relativistic Quantum Field Models},
  Annals Phys.\  {\bf 120}, 253-291 (1979).

\bibitem{FP1}
Paul-Emile Maing\'e, 
{\em Fixed point iterations coupled with relaxation factors and inertial effects},
 Nonlinear Analysis: Theory, Methods and Applications
{\bf 72-2}, 720-733 (2010).


\bibitem{FP2}
J.~Driesen, R.~Belmans, K.~Hameyer and J.~Fransen,
{\em Adaptive relaxation algorithms for thermo-electromagnetic FEM problems},
IEEE trans. on magnetics, {\bf 35-3}, 1622-1625 (1999).

\end{thebibliography}
\end{document}